\def\puncspace{\ifmmode\,\else{\ifcat.\C{\if.\C\else\if,\C\else\if?\C\else%
\if:\C\else\if;\C\else\if-\C\else\if)\C\else\if/\C\else\if]\C\else\if'\C%
\else\space\fi\fi\fi\fi\fi\fi\fi\fi\fi\fi}%
\else\if\empty\C\else\if\space\C\else\space\fi\fi\fi}\fi}
\def\SP{\let\\=\empty\futurelet\C\puncspace}

\def\h1{$h^{-1}$\SP}

\def\etal{{\it et al.\/}\ }

\def\eg{{\it e.g.\/}\rm,\ }
\def\lsim{~\rlap{$<$}{\lower 1.0ex\hbox{$\sim$}}}
\def\gsim{~\rlap{$>$}{\lower 1.0ex\hbox{$\sim$}}}
\def\void#1{{}}
\documentclass{aa}
\usepackage{graphicx}
\begin{document}

   \thesaurus{06     % A&A Section 6: Form. struct. and evolut. of stars
              (03.11.1;  % Cosmogony,
               16.06.1;  % Planets and satellites: general,
               19.06.1;  % Solar system: general,
               19.37.1;  % Stars: formation of,
               19.53.1;  % Stars: oscillations of,
               19.63.1)} % Stars: structure of.
   \title{ESO Imaging Survey}

   \subtitle{I. Preliminary Results}

\author { M. Nonino \inst{1,2} \and E. Bertin\inst{1,3,4} \and L. da 
Costa\inst{1} \and E. Deul\inst{1,3} \and T. Erben\inst{1,5} \and L. 
Olsen\inst{1,6} \and I. Prandoni\inst{1,7} \and M. Scodeggio\inst{1}
\and A. Wicenec\inst{1} \and  R. Wichmann\inst{1,8} \and C.
Benoist\inst{1,9} \and W. Freudling\inst{10} \and M.D.
Guarnieri\inst{1,11} \and I. Hook\inst{1} \and R. Hook\inst{10}  \and R.
Mendez\inst{1,12} \and S. Savaglio\inst{1}
 \and D. Silva\inst{1} \and R. Slijkhuis\inst{1,3}
}

\institute{
European Southern Observatory, Karl-Schwarzschild-Str. 2,
D--85748 Garching b. M\"unchen, Germany \and
Osservatorio Astronomico di Trieste, Via G.B. Tiepolo 11, I-31144
Trieste, Italy \and
Leiden Observatory, P.O. Box 9513, 2300 RA Leiden, The Netherlands \and
Institut d'Astrophysique de Paris, 98bis Bd Arago, 75014 Paris, France \and
Max-Planck Institut f\"ur Astrophysik, Postfach 1523 D-85748,  Garching b. 
M\"unchen, Germany \and
Astronomisk Observatorium, Juliane Maries Vej 30, DK-2100 Copenhagen, 
Denmark \and
Istituto di Radioastronomia del CNR, Via Gobetti 101, 40129 Bologna, Italy \and
Landensternwarte Heidelberg-K\"onigstuhl, D-69117, Heidelberg, Germany \and
DAEC, Observatoire de Paris-Meudon, 5 Pl. J. Janssen, 92195 Meudon Cedex, 
France \and
Space Telescope -- European Coordinating Facility, Karl-Schwarzschild-Str. 2, 
D--85748 Garching b. M\"unchen, Germany \and
Osservatorio Astronomico di Pino Torinese, Strada Osservatorio 20, I-10025 
Torino, Italy \and
Cerro Tololo Inter-American Observatory, Casilla 603, La Serena, Chile
}

%   \author{EIS Team
%          \inst{1}
%          \and
%         C. Ptolemy\inst{2}\fnmsep\thanks{Just to show the usage
%          of the elements in the author field}
%          }

   \offprints{M. Nonino}

%   \institute{European Southern Observatorry, Karl-Schwazschild-Str. 2
%   Garching bei M\'unchen, Germany\\
%              email: 
%         \and
%             University of Alexandria, Department of Geography\\
%             email: c.ptolemy@hipparch.uheaven.space
%             \thanks{The university of heaven temporarily does not
%                     accept e-mails}
%             }

%   \date{Received September 15, 1996; accepted March 16, 1997}

   \maketitle
    
   \today

   \begin{abstract}

This paper presents the preliminary results of the ESO Imaging Survey
(EIS), a public survey being carried out by ESO and member states, in
preparation for the VLT first-light. The survey goals, organization,
strategy and observations are discussed and an overview is given of
the survey pipeline developed to handle EIS data and produce object
catalogs. A report is presented on moderately deep I-band observations
obtained in the first of four patches surveyed, covering a region of
3.2 square degrees centered at $\alpha
\sim 22^h 40^m$ and $\delta =-40^\circ$. The products available to 
the community, including pixel maps (with astrometric and photometric
calibrations) and the corresponding object catalogs, are also
described. In order to evaluate the quality of the data, preliminary
estimates are presented for the star and galaxy number counts, and for
the angular two-point correlation function obtained from the available
data.  The present work is meant as a preview of the final release of
the EIS data that will become available later this year.

   \end{abstract}

%
%________________________________________________________________

\section{Introduction}

With the advent of very large telescopes, such as the VLT, a largely
unexplored domain of the universe becomes accessible to observations
which may dramatically enhance on our understanding of different
physical phenomena, in particular the origin and evolution of galaxies
and large scale structures.  In the next few years a wide array of 8-m
telescopes will become available world-wide.  Among these, the
European VLT project is particularly striking because of its four 8-m
telescopes and an impressive array of complementary
instrumentation. Viewed as a unit, the VLT provides great flexibility
by combining complementarity for certain programs with multiplexing
capabilities for others. First-light for the VLT is scheduled for May
1998, with regular science operation starting in April 1999.

In order to take full-advantage of the VLT from the start of its
operation, ESO and its Observing Programmes Committee (OPC) decided to
coordinate an imaging survey to provide candidate targets well-suited
to the first set of VLT instruments. The ESO Imaging Survey (EIS) has
been conceived as a collaborative effort between ESO and astronomers
in its member states. Following the recommendation of the OPC, the
survey has been overseen by a Working Group (WG).  The EIS WG is
composed of leading experts in different fields and has the
responsibility of defining the survey science goals and strategy, and
monitor its progress. In order to carry out the survey a dedicated
team was assembled, starting March 1997. To stimulate cooperation
between ESO and the astronomical community of the member states, EIS
has sponsored the participation of experts as well as students and
post-docs from the community in the development of software,
observations and data reduction.

As described by Renzini \& da Costa (1997) (see also
``http://www.eso.org/eis''), EIS consists of two parts: EIS-wide to
search for rare objects (\eg distant clusters and quasars) and
EIS-deep to define samples of high-redshift galaxies.  These science
goals were chosen to match as well as possible the capabilities of the
first VLT instruments, FORS, ISAAC and UVES.  EIS is also an
essential first step in the long-term effort, currently underway at
ESO, to provide adequate imaging capabilities in support of VLT
science (Renzini 1998). The investment made in EIS will be carried
over to a Pilot Survey utilizing the ESO/MPIA 2.2m telescope at La
Silla, with its new wide-field camera. This Pilot Survey, which will
follow the model of EIS, has been recommended by the EIS WG and is
being submitted to the OPC.

The goal of this paper is to describe the characteristics of I-band
observations carried out in the fall of 1997 over a region of 3.2
square degrees (EIS patch A, da Costa \etal 1998a) and of the
corresponding data products, in the form of calibrated images and
single frame catalogs.  These products have been made publicly
available through the ESO Science Archive, as a first step towards the
full distribution of the EIS data.  The purpose of the present release
is also to provide potential users with a preview of the data, which
may help them in the preparation of VLT proposals, and to encourage
the community to provide constructive comments for the final
release. It is important to emphasize that due to time limitations the
results presented here should be viewed as preliminary and
improvements are expected to be made before the final release of the
EIS data later this year.

In section 2, a brief description is presented of the criteria adopted
in the field selection, the strategy of observations and the
characteristics of the data in patch A already completed. It also
describes the filters used, the definition of the EIS magnitude system
and its relation to other systems, and the data used for the
photometric calibration of the survey. In section 3, a brief
description of the data reduction pipeline is presented, followed in
section 4 by a description of the data products made publicly
available in this preliminary data release. In section 5, the
algorithms used to detect and classify objects, and the information
available in the catalogs being distributed are described. Preliminary
results from a scientific evaluation of the data is presented in
section 6. In section 7, future plans are presented, followed in
section 8 by a brief summary.

\section{The Survey}

\subsection{Goals}

EIS-wide is a relatively wide-angle survey of four pre-selected
patches of sky, 6 square degrees each, spanning the right ascension
range $22^h < \alpha < 9^h$. The main science goals of EIS-wide are
the search for distant clusters and quasars. To achieve these goals
the original proposal envisioned the observation of 24 square-degrees
in V and I, 6 square degrees in B over one of the patches, and 2
square degrees in U in a region near the South Galactic Pole.

Because of the slow start of the survey due to the unusually bad
weather caused by El Ni\~no, which dramatically affected the
observations in the period of July-November 1997, some of these goals
had to be reassessed by the WG (da Costa \etal 1998a). It was decided
to limit the observations to the I-band, except for patch B, the
region close to the South Galactic Pole, where observations were
conducted in B, V and I over 1.5 square degrees. The current status of
the observations for EIS-wide is summarized in Table~\ref{tab:status},
where the J2000 centers of the actually surveyed patches and the area
covered in the different bands are given.

\begin{table}
\caption{Current Sky Coverage}
\label{tab:status}
\begin {tabular}{lcccccc}
\hline
Patch   & $\alpha$ & $\delta$& B  & V   & I  \\
\hline
A       & 22:42:54  &   -39:57:32 & -   & 1.2  & 3.2 \\
B       &  00:49:25 &   -29:35:34 & 1.5  & 1.5  & 1.6  \\
C       &  05:38:24  &  -23:51:00  &  -   &  -   & 6.0  \\
D       &   09:51:36 &   -21:00:00 &-   &  -   & 6.0 \\
\hline
        &    -       &      -      & 1.5  & 2.7  & 16.8  \\
\hline
\end{tabular}
\end{table}

EIS-deep is a multicolor survey in four optical and two infrared bands
covering 75 arcmin$^2$ of the HST/Hubble Deep Field South (HDFS),
including the WFPC2, STIS and NICMOS fields, and a region of
100~arcmin$^2$ in the direction of the southern hemisphere counterpart
of the Lockman Hole, to produce samples with photometric redshifts to
find U-dropout candidates and galaxies in the redshift range $ 1 < z <
2$. Observations for EIS-deep will start in August 1998 and therefore
this part of the survey is not discussed in the present paper.

\subsection {Field Selection}

The four EIS-wide patches were selected to have, in general, low
optical extinction, low FIR ($\lambda=$ 100 $\mu$, $A_V < 0.05$)
emission and low HI column density ($\sim 2 \times 10^{20}$cm$^{-2}$). They
were also examined to guarantee that they would not include very
bright stars or nearby clusters of galaxies. Preference was also given
to fields that would overlap with other interesting datasets,
especially recently completed or ongoing wide-angle radio surveys (\eg
NVSS, Westerbork in the Southern Hemisphere, Australia Telescope ESO
Slice Project). The fields were also chosen to cover a range of
galactic latitudes of possible interest for galactic studies.

\subsection {EIS Filters}
\label{filter}

\begin{figure}
\resizebox{\hsize}{!}{\includegraphics{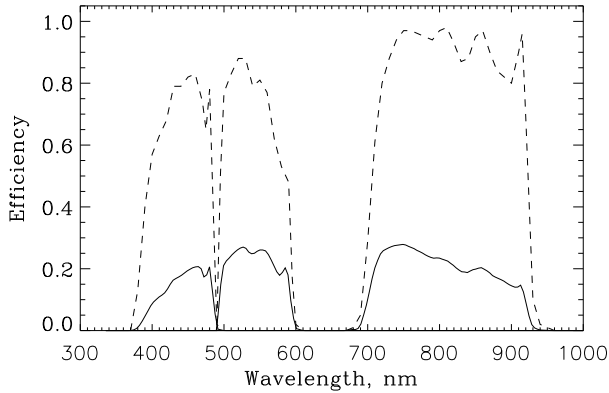}}               
\caption{Transmission curves for the EIS filters (dashed lines)  and
total system throughput including the contribution from the filters,
telescope and camera optics, and the detector (solid lines).}
\label{fig:filters}
\end{figure}

Since the primary consideration in the selection of the filters was
the desire for depth, the observations were conducted using
wide-passband filters. EIS uses a special set of $BVI$ filters (ESO
WB430\#795, WB539\#796, WB829\-\#797), which were designed to have
higher transmission than the BVI$_C$ passbands. The transmission
curves for the filters and the full response of the NTT-EMMI red system
with the EIS filters are shown in Figure~\ref{fig:filters}, and can be
retrieved in electronic form from the World Wide Web at
``http://www.eso.org/eis/eis\_filters.html''.

While the effective wavelengths of these filters are close to those of
the Johnson-Cousins $BVI_c$ filters, their passbands are broader and
have sharper cutoffs. The measured passbands of these filters have been
used to derive synthetic photometry using the Gunn \& Stryker (1983)
catalog of spectrophotometric scans of main-sequence and giant stars. 
For WB430\#795 (B band) the throughput was determined to be 0.42~mag
higher than Johnson B (at $B-V=0$), and for WB829\#797 0.44~mag higher
than Cousins $I_c$ (at $V-I_c$=0).  However, the WB539\#796 filter
turned out to have a throughput slightly lower (0.18~mag) than Johnson
V.

\subsection{Survey Strategy}

The observations for EIS-wide started in July 1997 and are being
conducted using the EMMI camera (D'Odorico 1990) mounted on the 3.5-m
New Technology Telescope (NTT) at La Silla. The EMMI red channel is
equipped with a Tektronix $2046 \times 2046$ chip with a pixel size of
0.266 arcsec and an effective field-of-view of about $9' \times 8.5'$,
because of the strong vignetting at the top and bottom parts of the
CCD. In order to cover a large area of the sky the observations are
being conducted using a sequence of 150 sec exposures shifted by half
the size of an EMMI-field both in right ascension and in
declination. This leads to an image mosaic whereby each position in
the sky is observed twice for a total integration time of 300 sec,
except at the edges of the patch.  The observations have been made
using the new Data Flow System installed on the NTT which requires the
preparation of observational blocks (OBs) that control the sequence of
the observations.  Using the Phase-2 Proposal Preparation (P2PP)
software, a typical EIS OB consists of a sequence of ten frames. The
telescope presets to the target position specified in the OB, exposes
for 150 sec and thereafter offsets from this position according to a
specified pattern: a shift east and north followed by a shift east and
south. The EIS mosaic thus consists of frames with significant
overlaps (a quarter of an EMMI frame). The easiest way of visualizing
the geometry of the EIS mosaic is to picture two independent sets of
tiles (referred to as $P_{i,j}$, where i refers to the row along right
ascension and j to columns in declination) forming a contiguous grid
(normally referred to as odd and even depending on the value of j, and
P is the patch name A-D) superposed and shifted in right ascension and
declination by half the width of an EMMI frame.  To ensure continuous
coverage, adjacent odd/even frames have a small overlap at the edges
($\sim 20$ arcsec).  Therefore, a small fraction of the surveyed area
may be covered by more than two frames. Such a mosaic ensures good
astrometry, relative photometry and the satisfactory removal of cosmic
ray hits and other artifacts.

\subsection {Observations}

\begin{figure}
\resizebox{\hsize}{!}{\includegraphics{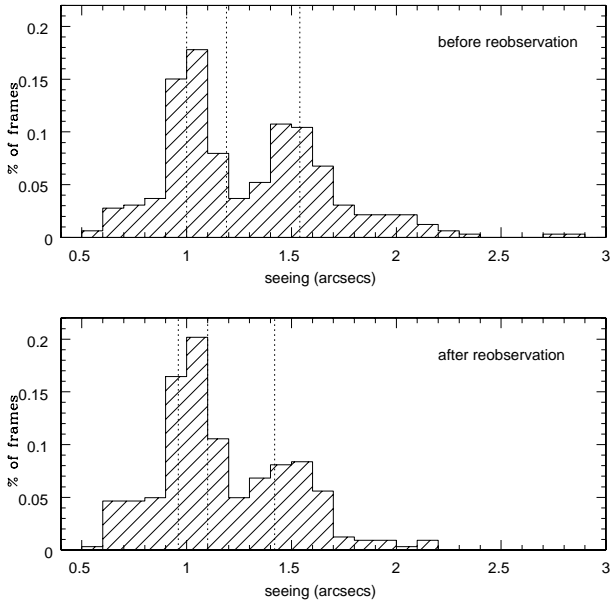}}               
\caption{ Seeing distribution for patch A obtained from 
all observed frames (top panel) and only from the frames actually
accepted for the survey (bottom panel). Vertical lines refer to 25, 50
and 75 percentiles of the distributions.}
\label{fig:seeing}
\end{figure}

Because of bad weather conditions, for patch A it was only possible to
cover 3.2 square degrees in the I-band and 1.2 square degrees in the
V-band.  These observations were obtained in six different runs in the
fall of 1997. The observations were carried out in standard visitor
mode and data were taken in less than ideal conditions. A total of 400
science frames were obtained in I-band for patch A with the seeing
varying from about 0.6 arcsec to over 2 arcsec. Regions observed under
poor conditions were re-observed to maintain some degree of uniformity
in the depth of the survey. In Figure~\ref{fig:seeing} the seeing
distribution of all frames obtained in patch A (top panel) and of the
frames actually accepted for the survey (bottom panel) are compared.
This gives an idea of the area which required re-observations ($\lsim
1$ square degree) and the impact that these had on the early progress
of the survey.  The median seeing for all frames is about 1.2 arcsec,
while for the accepted frames it is about 1.1 arcsec. More
importantly, the fraction of frames with seeing greater than 1.5
arcsec was greatly reduced. Note that the observing conditions have
varied considerably within a run and from run to run during the period
of observations of patch A, with only a small number of photometric
nights. Therefore, depending on the application it may be necessary to
further filter the object catalogs according to the characteristics of
the data (Section~\ref{field}). While this is trivial when dealing
with single frames, proper pruning of the data requires more
sophisticated tools when considering catalogs produced from the
coadded image (\eg da Costa
\etal 1998a).

\begin{figure}
\resizebox{\hsize}{!}{\includegraphics{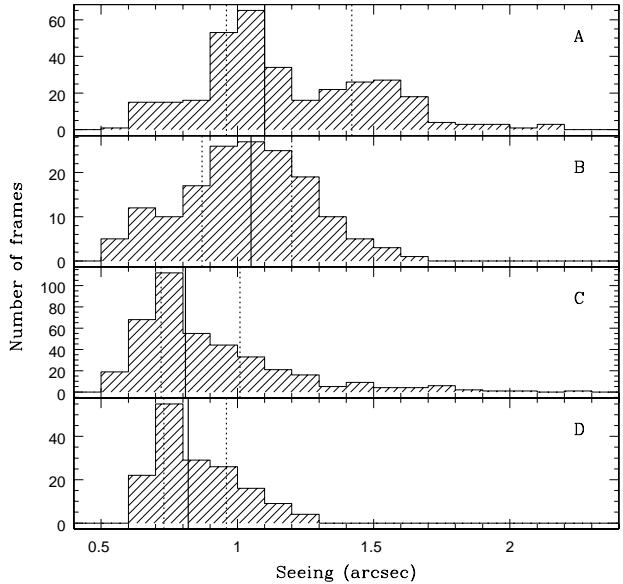}}               
\caption{ Seeing distribution for all the EIS patches. The plot
includes all the data taken in 1997. Note the great improvement of the
seeing distribution, especially for patches C and D.}
\label{fig:seeingall}
\end{figure}

It is worth emphasizing that the data for patch A are by far the
worst. This can be seen in Figure~\ref{fig:seeingall}, where the
seeing distribution for patch A is compared with that of the other
patches.

During EIS nights photometric and spectrophotometric standards are
regularly observed ({\cite{landolt1}, Baldwin \& Stone 1984,
\cite{landolt2}). Whenever possible, photometric solutions
 are derived to evaluate the quality of the night and to determine
absolute zero-points for the tiles observed during photometric nights.

The photometric quality of the nights is also being assessed from the
observations conducted by the Geneva Observatory at the 0.7m Swiss
telescope, which regularly monitors the extinction coefficients at La
Silla. The information from the Swiss telescope is stored in a
calibration database to provide the necessary reference with the
survey nights. Whenever this information is available, images taken
during photometric nights are flagged, and this information is used in
the absolute photometric calibration of the patch
(section~\ref{photom}).

\subsection {EIS Magnitude System}

The zero-points of the EIS magnitude system have been defined to give
the same B, V, I magnitudes as in the Landolt system for stellar
objects with $(B-V)=(V-I_c)=~0$.  In other words, EIS magnitudes are
by definition equal to the magnitudes in the Johnson-Cousins system
for zero-color stars. 

In Figure~\ref{fig:color} we show the observed transformation between
the EIS system and Johnson-Cousins for the I-band, as a function of
color.  The data points are based on all of the reduced observations
of standard stars currently available (run 1-7). From a linear fit to
the data points the color term between the EIS and Counsins system is
found to be small ($0.014 \pm 0.004$). Note, however, that because of
the limited amount of photometric nights up to run 7, color terms have
not yet been taken into account in the photometric solutions, and the
current I-band zero-point of the EIS system may be subject to small
changes ($\sim~0.01$mag) at a later date, when more data becomes
available in the final release.

\begin{figure}[ht]
\resizebox{\hsize}{!}{\includegraphics{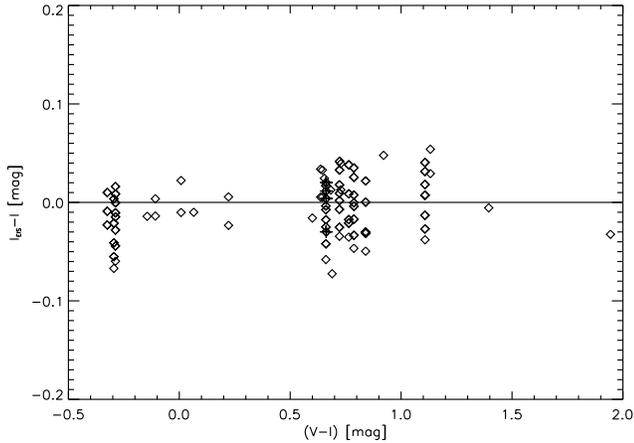}}               
\caption{Relation between the EIS and Johnson-Cousins system as a
function of color.  Shown are all the standard stars observed under
photometric conditions in the period July 97-January 98.}
\label{fig:color}
\end{figure}

\void{
Based on both the observation of standards with the NTT and the Swiss
information, the number of tiles in patch A observed under photometric
conditions is estimated to be about 10\% of the total number of
accepted tiles. It is important to point out that the conditions were
much better for all the other EIS patches.}

\subsection {External Data}
\label{zero}

In order to provide additional constraints on the absolute calibration
of the survey, data were obtained under photometric conditions from
other telescopes at La Silla. Observations were conducted at the 2.2m
telescope (4 half-nights, out of which one half-night under
photometric conditions) and one night at the 0.9m Dutch telescope.

The observations with the 2.2m telescope were carried out using EFOSC2
with a 2048x2048 CCD chip, with a pixel size of 0.27 arcsec and a
seeing of about 1.5 arcsec. The filters used in these observations
were ESO\#583, ESO\#584, ESO\#618. EFOSC2 has a field of view $8.3'
\times 7.7'$. During observations of the only photometric night,
 standard stars and frames over the original area of patch A were
obtained in B, V and I.  Unfortunately, only two I-band images overlap
with the surveyed area, as shown in Figure~\ref{fig:overlaps}. The
limiting magnitude of the frames is $I \sim 20$. Over 100
objects per frame were found in common with the overlapping survey
field.

\begin{figure}[ht]
\resizebox{\hsize}{!}{\includegraphics{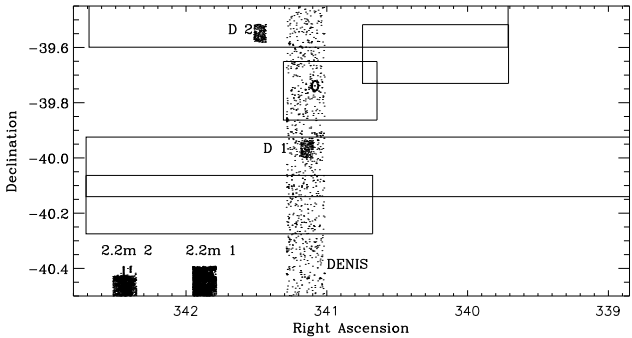}}               
\caption{Distribution of frames obtained  at the 2.2m and
the 0.9m Dutch (D) telescopes at La Silla overlapping the surveyed
region of patch A. Also shown is a part of a DENIS strip that crosses
the field. The open rectangles represent regions containing EIS tiles
observed under photometric conditions.}
\label{fig:overlaps}
\end{figure}

The observations with the 0.9m Dutch telescope were done in one
photometric night using TK512CB (a 512x512 CCD chip, with a pixel size
of 0.47 arcsec) at a seeing of about 1.5 arcsec. The field of view
approximates one quarter of an EIS survey tile, i.e. $3.1' \times
4.0'$. The filters used were ESO\#419, ESO\#420, and
ESO\#465. The limiting magnitude of the science frames is $I \sim
22$. Approximately 200 objects per frame were found in common with the
overlapping EIS survey tiles for the range $ 14 < I < 22$ to define
the zero-point.

Images from both telescopes were processed in a standard way using
IRAF. One major problem was the severe fringing observed in the I-band
frames from the 2.2m.  The fringing was removed by creating a
combination of the 300s science exposures and using IRAF's mkfringcor
task, but preliminary results that this correction may have
affected the measurement of faint objects.

In addition to these scattered fields, a reduced strip of I-band data
from the DENIS survey (Epchtein~\etal 1996) is also
available. Unfortunately, close inspection of the data showed that the
strip was observed in non-\-photometric conditions.

Figure~\ref{fig:overlaps} shows the overlap of the external data with
the surveyed region.  Also shown are the regions covered by EIS tiles
observed in photometric conditions.

\section {Data Reduction}

\subsection{ Overview of the  EIS Pipeline}

An integral part of the EIS project has been the development of an
automated pipeline to handle and reduce the large volume of data
generated by EIS. The pipeline consists of different modules built
from preexisting software consisting of: 1) standard IRAF tools for
the initial processing of each input image and preparation of
superflats; 2) the Leiden Data Analysis Center (LDAC) software,
developed for the DENIS (\cite{denis}) project to perform photometric
and astrometric calibrations; 3) the SExtractor object detection and
classification code (\cite{bertin}); 4) the ``drizzle'' image
coaddition software (Fruchter \& Hook 1997, Hook \& Fruchter 1997),
originally developed for HST, to create coadded output images from the
many, overlapping, input frames.

A major aim of the EIS software is to handle the generic problem posed
by the building up of a mosaic of overlapping images, with varying
characteristics, and the extraction of information from the resulting
inhomogeneous coadded frames. This has required significant changes in
the preexisting software. To illustrate the power of the tool being
developed a brief description of the basic idea behind the pipeline is
necessary.  For each input frame a weight map, which contains
information about the noise properties of the frame, and a flag map,
which contains information about the pixels that should be masked such
as bad pixels and likely cosmic ray hits, are produced.  After
background subtraction and astrometric and relative photometric
calibration, each input frame is mapped to a flux-preserving conic
equal-area projection grid, chosen to minimize distortion in area and
shape of objects across the relatively large EIS patch. The flux of
each pixel of the input frame is redistributed in the superimage and
coadded according to weight and flags of the input frames contributing
to the same region of the coadded image.  Therefore, the coaddition is
carried out on a pixel-by-pixel basis.  It is clear that in the
process the information about the individual input frames is lost and
in order to trace them back an associated context (domain) map is
created, providing the required cross-reference between the object and
the input frames that have contributed to its final flux. Another
important output is the combined weight map which provides the
required information to the object detection algorithm to adapt the
threshold of source extraction to the noise properties of the context
being analyzed. SExtractor has been extensively modified for EIS and
its new version incorporates this adaptive thresholding (new
SExtractor documentation and software are available at
``http://www.eso.org/eis/eis\_soft.html''). For a survey such as EIS,
being carried out in visitor mode with varying seeing conditions, this
cross-reference is essential as it may not be possible to easily
characterize the PSF in the final coadded image, a problem which in
turn affects the galaxy/star classification algorithm.

In the present paper the main focus is on the processing and the
object catalogs extracted from the individual frames that make up the
mosaic and which provide a full contiguous coverage of patch A with
well defined characteristics. A detailed discussion of coadded images
and the catalogs extracted from them will be presented when these data
become public later this year.

\void {It is clear that working with
the source catalogs extracted from the coadded images will be similar
except that the data quality control has to be done using the
contexts, which represent virtual frames, instead of single frames.}

\subsection {Retrieving Raw Data}

EIS utilizes the observational and the technical capabilities of the
refurbished NTT and of the ESO Data Flow System (DFS), from the
preparation of the observations to the final archiving of the data. DFS
represents an important tool for large observational programs.
The ESO Archive is interfaced with EIS pipeline at both ends: it
supplies the observed raw data and collects the output catalogs and
reduced images.

\void{The data delivery of the raw data to the main reduction platform of
EIS has been carried out by implementing a 'data subscriber', as used
in several places throughout the DFS. A customized setup of this
subscriber ensures delivery of EIS relevant data only.}

In the course of delivering the data to EIS, the raw data is also
archived in the ESO NTT Archive. The headers of the data delivered to
EIS had to be adjusted in a variety of ways to meet the requirements
of the EIS pipeline. Since standard ESO FITS-headers contain a wealth
of instrumental and observational parameters in special ESO keywords,
translation of some of these keywords to user defined keywords is a
standard tool the Archive offers. Header translation was executed
automatically right after the transfer of every single file. Moreover
the files were renamed and sorted into subdirectories to reflect the
nature of the frames (calibration, science and test frames). Filenames
have been constructed which ensured both uniqueness and a meaningful
description of the observed EIS tile, filter and exposure. Most of the
important parameters of all transferred files are ingested into
several database tables, briefly described below, which serve as a
pool of information used during the execution of the pipeline. The
database also provides all the information necessary to characterize
the observations.

\subsection {Frame Processing}
\label {frame}

The EMMI frames have been read in a two-port readout mode. The assumed
gain difference between the two amplifiers is  about 10\% (2.4
e$^{-}$/ADU, 2.16 e$^{-}$/ADU), with a readout noise of 5.49 e$^{-}$
and 5.81e$^{-}$, as reported by the NTT team. Slight variations of this
value have been detected from run to run. Currently, standard IRAF
tools are used to remove the instrumental signature from each
frame. To handle the dual-port readout, pre-scan corrections are
applied for each half-frame using the {\em xccdred} package of IRAF.

A master bias for each run is created by median combining all bias
frames, typically 10 each day, using the 3-$\sigma$ clipping option.
The effective area used in this calculation is from columns 800 to 2000
(along the x-axis) and rows from 100 to 1900 (along the y-axis), which
avoids a bad column visible at the upper part of the chip and
vignetted regions on top and bottom of the image. The same procedure
is adopted for the dome flats. About 10 dome flats for each filter
were used with about 15,000 ADUs. Skyflats were obtained
at the beginning and end of each night using an appropriate calibration
template to account for the variation of sky brightness in each band
so that about five skyflats per night were obtained with 40,000 ADUs.
These bright sky exposures were monitored automatically to reject
those frames which could have been saturated or that had low S/N. The
skyflats were combined using a median filter on a run basis and
then median-filtered within a $15\times 15$ pixel box.

For each science frame a pre-scan correction is made and the frames
are trimmed making the usable area of the chip from row 21 to 2066
(x-axis) and from column 1 to 2007 (y-axis). Note that this procedure
does not completely remove the vignetted region of the frame or a
coating defect visible on the chip. As  will be seen later a suitable
mask is defined to handle these regions.  After trimming, the combined
bias is subtracted and the frames are flat-fielded using the dome
flats, and an illumination correction is applied using the combined
skyflats.

After each survey frame has been corrected for these instrumental
effects, a quick inspection is carried out by eye. A single EIS
keyword is set in the header to flag those frames which contain
satellite tracks, bright star(s), background gradients due to nearby
bright objects, or those useless for scientific purposes (\eg those
affected by motions of the EMMI rotator or by glitches in the tracking
leading to double or even triple images). Flagged frames are then
rejected during the creation of the supersky flat, which is obtained
from the combination of all suitable science frames using a median
filter and 3$\sigma$ clipping. The superflat is created on a run basis
and is smoothed using a running box 15 x 15 pixels in size.

During visual inspection masks are also created and saved for use by
the pipeline (section~\ref{weight}) to mark regions affected by bright
stars just outside the frame, some satellite tracks or cosmic rays in
the form of long streaks in the frame. This can be done by taking
advantage of features recently implemented in the SkyCat display tool
(Brighton 1998, see also ``http://www.eso.org/eis/eis\_soft.html'').

Finally, the superflat is applied to both the survey and standard star
frames taken during the night. The background images generated by
SExtractor, prior to the source extraction (minibacks), have been used
to estimate the homogeneity of the resulting science images. In
general, the flatness of the images is $\lsim 0.2\%$, except for two
runs for which larger values (up to $\sim 1.4\%$) are found (Slijkhuis
\etal 1998).  A major contribution to the background
residual is probably due to variations in the relative gain of the two
readout ports, which will be investigated further.

Note that the flat-fielding of EIS images is done in {\em surface
brightness}, not in {\em flux}. Variations of the pixel scale over the
field may cause a drift of the magnitudes, especially at the edges of
the frames.  However, distortions lead to a variation of pixel-scale
which has been estimated from the astrometric solution
(Section~\ref{astrom}) to be $\lsim 0.5$\%. This translates to a
photometric drift $\lsim 0.01$~mag, over the field, which has not been
corrected for in the present release.

\subsection {Processing Standard Stars}

Frames for standard stars are also processed automatically through a
parallel branch of the pipeline fine-tuned to process standard star
fields.  Reference catalogs for all the Landolt fields are available
and are used to pair objects and identify the standard stars. Aperture
photometry, using Landolt apertures, is carried out and extinction
coefficients and zero-points are computed and stored in the
calibration database together with other observational parameters.
Plots are also produced to ease the task of identifying photometric
nights. The automatic process has been checked against reductions
carried out manually with IRAF tasks, yielding consistent results.

\void{
Using this automatic processing several errors in the coordinates
of Landolt standards have been found an reported.}

\subsection {Survey monitoring and Quality Control}
\label{ksb}

An integral part of the survey pipeline is the automatic production of
reports for the monitoring of the data and the data reduction, and for
diagnosing the different tasks of the EIS pipeline.  These reports produce
several plots that are interfaced with the WEB, for easy access by the
EIS team. They can also be retrieved at a later time from information
available in the EIS database (section~\ref{database}).

From the raw data retrieved from the archive a set of plots is
produced which provides information on: 1) the time-sequence of
observations (which has helped optimize the use of the DFS, and
monitor the efficiency of the observations); 2) the performance of the
pointing model; 3) the continuity of the coverage of the patch (by
monitoring the overlap between tiles); and 4) the observed tiles and
repeated observations.

After processing the data another set of plots is produced which show
the seeing as measured on the images, the number counts at a fixed
limiting magnitude, the limiting magnitude, defined as the $5\sigma$
detection threshold for a point source, and the $1\sigma$ limiting
isophote (mag/arcsec$^2$).  As an illustration
Figure~\ref{fig:limsbmag} shows the distribution of the limiting
isophote and magnitude for the accepted tiles of patch A, and
Figures~\ref{fig:seeingmap} and~\ref{fig:limmagmap} show the
two-dimensional distribution of the seeing and limiting
magnitude. These plots are useful to guide the selection of regions
for different types of analysis.  Diagnostic plots are also produced
after the astrometric calibration of the frames, and before and after
the calculation of the relative and the absolute photometric
calibration.

\begin{figure}
\resizebox{\hsize}{!}{\includegraphics{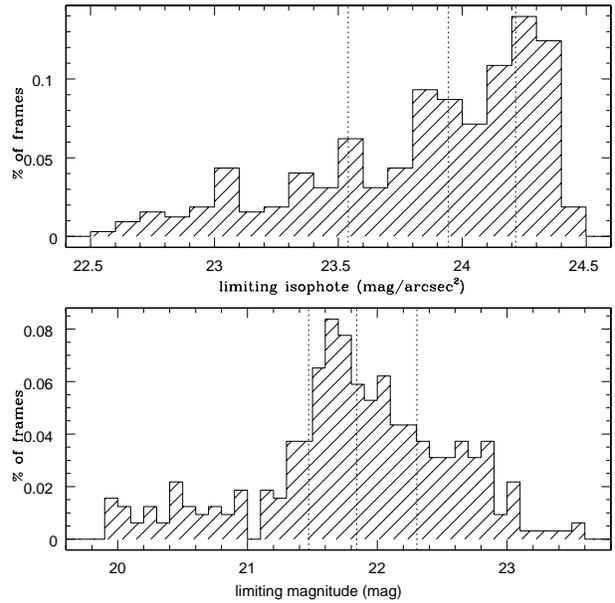}}               
\caption{  Limiting isophote (top panel) and limiting magnitude 
(bottom panel) distributions from the patch A frames actually accepted
 for the survey.  Vertical lines refer to 25, 50 and 75 percentiles of
 the distributions.}
\label{fig:limsbmag}
\end{figure}

Based on this information the observing plans for subsequent
runs are reviewed. In the case of patch A, originally observed
under poor conditions, an attempt was made to improve the quality of
the data as discussed earlier (see Figure~\ref{fig:seeing}).

\begin{figure}[ht]
\resizebox{\hsize}{!}{\includegraphics{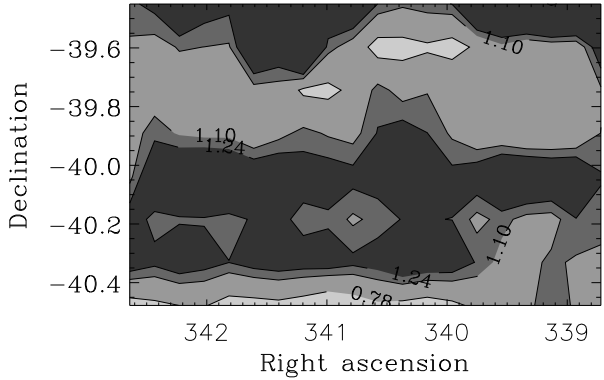}}               
\caption{Two-dimensional distribution of the seeing as measured on the
I-band images for patch A for all the accepted even frames. Contours
refer to 25, 50 and 75 percentiles of the distribution.}
\label{fig:seeingmap}
\end{figure}

\begin{figure}[ht]
\resizebox{\hsize}{!}{\includegraphics{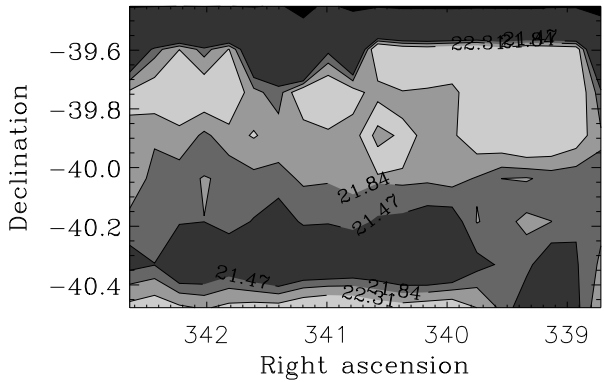}}               
\caption{Two-dimensional distribution of the computed 5$\sigma$
limiting magnitude for point sources estimated from the accepted
even frames for patch A.  Contours refer to 25, 50 and 75 percentiles of the
distribution.}
\label{fig:limmagmap}
\end{figure}  
%\subsubsection {Image Quality and EMMI PSF}

The image quality of the survey frames has also been monitored by
computing the size and anisotropy of the PSF for suitably chosen stars
covering the survey frames.  While this in principle could be done
directly from the catalogs produced by the pipeline, the second order
moments of the brightness distribution produced by the current version
of SExtractor are sensitive to noise which affect the measurement of
object shapes.  Therefore, to monitor the shape of the PSF the
software developed by Kaiser, Squires and Broadhurst (1995) (hereafter
KSB) has been implemented. It computes the shape of objects by using
appropriate weights in the calculation of the second order moments.
Comparison between the results of the two algorithms shows that, while
both lead to comparable results, the KSB software provides more robust
results than SExtractor (Figure~\ref{ksbsexcmp}). It is, however,
considerably slower than SExtractor, and for the time being it is run
in parallel to the main pipeline, just for diagnostic purposes.

\begin{figure*}[ht]
\includegraphics[width=12cm,height=4cm,angle=0.]{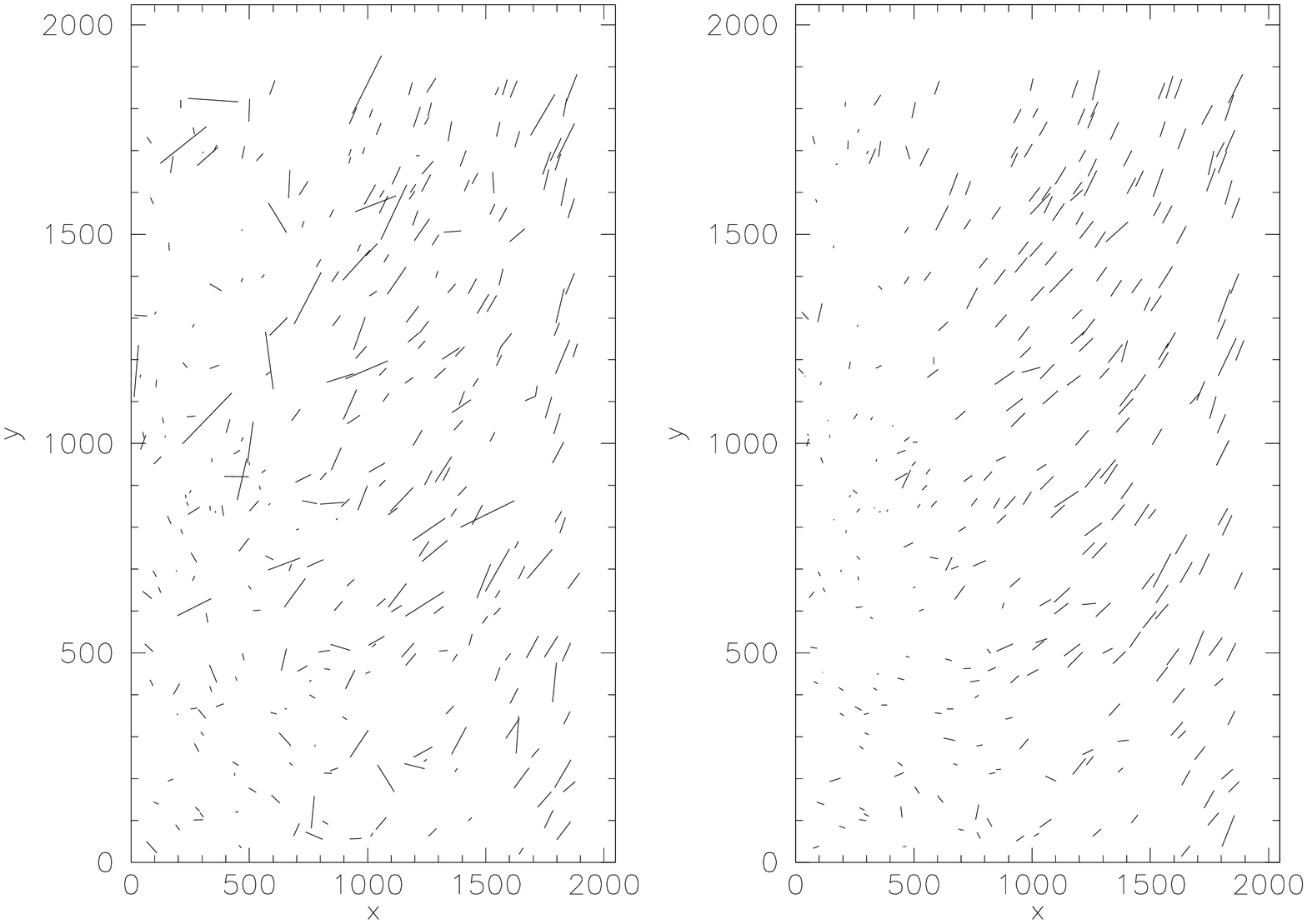}
\hfill
\parbox[b]{55mm}{
\caption{Comparison of SExtractor and KSB shapes for stellar objects:
  The left panel shows the anisotropy structure as measured by
  SExtractor, while the right panel as measured by the KSB algorithm
  for the same stars.}  \label{ksbsexcmp} }
\end{figure*}

The typical structure of the anisotropy of the EMMI PSF is shown in
the upper left panel of Figure~\ref{correct}, which displays the
polarization vector for stellar images in a frame with a seeing of
about 1 arcsec. As it can be seen the anisotropy shows a complex
structure and has a mean amplitude of $\sim 6\%$ (lower left
panel). However, the variation of the anisotropy is well represented
by a second-order polynomial. Application of this correction leads to
the small random residuals (rms $\sim 2$\%) shown in the right panels
of Figure~\ref{correct}. Tests have also shown that the number of
stars in the images of patch A is on average $\sim 40$. This number is
sufficient to allow this correction to be computed even for typical
survey frames.

\begin{figure*}[ht]
\includegraphics[width=12cm,height=12cm,angle=0.]{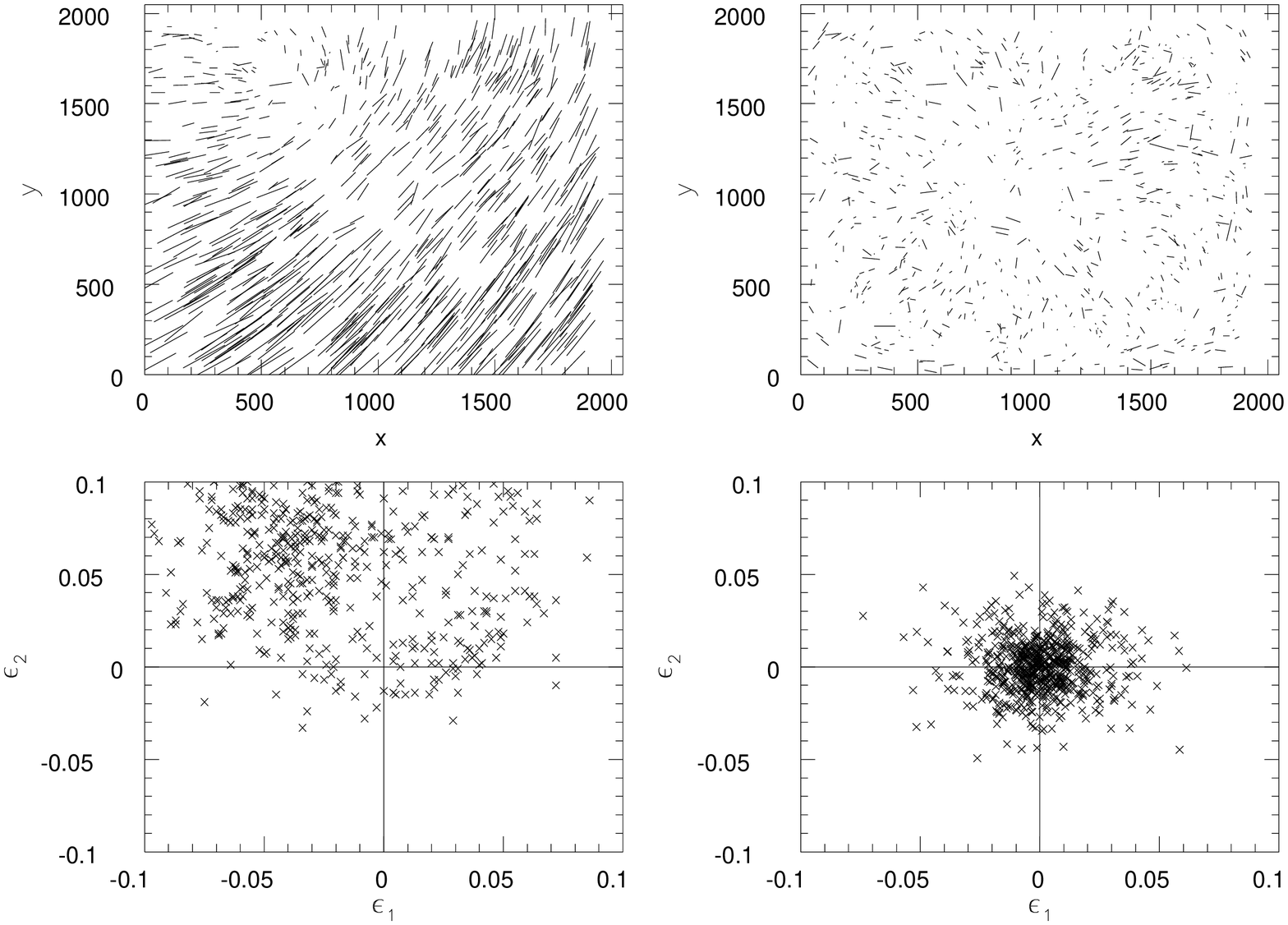}
\hfill
\parbox[b]{55mm}{
\caption{A typical pattern of the PSF anisotropy (upper left) and
components of the polarization vector (lower left) for an EMMI I-band
frame (1 arcsec seeing). Also shown are the spatial distribution
(upper right) and components of the polarization vector (lower right)
after polynomial correction, as described in the text.
}  \label{correct} }
\end{figure*}    

More importantly, the EMMI images show a systematic increase in the
size of the PSF along the y-direction of the CCD, while none is seen
in the x-direction (Figure~\ref{PSFsize}). The difference
between the size in the lower part and the upper part of the images is
typically $10\%$  which, in principle, can
affect the star/galaxy classification algorithm. This effect was
caused by the misalignment between the primary and secondary NTT
mirrors. Recently, the mirrors have been realigned and a great
improvement of the image quality is expected for the observations of
patch D.

\begin{figure*}[ht]
\includegraphics[width=12cm,height=6.0cm,angle=0.]{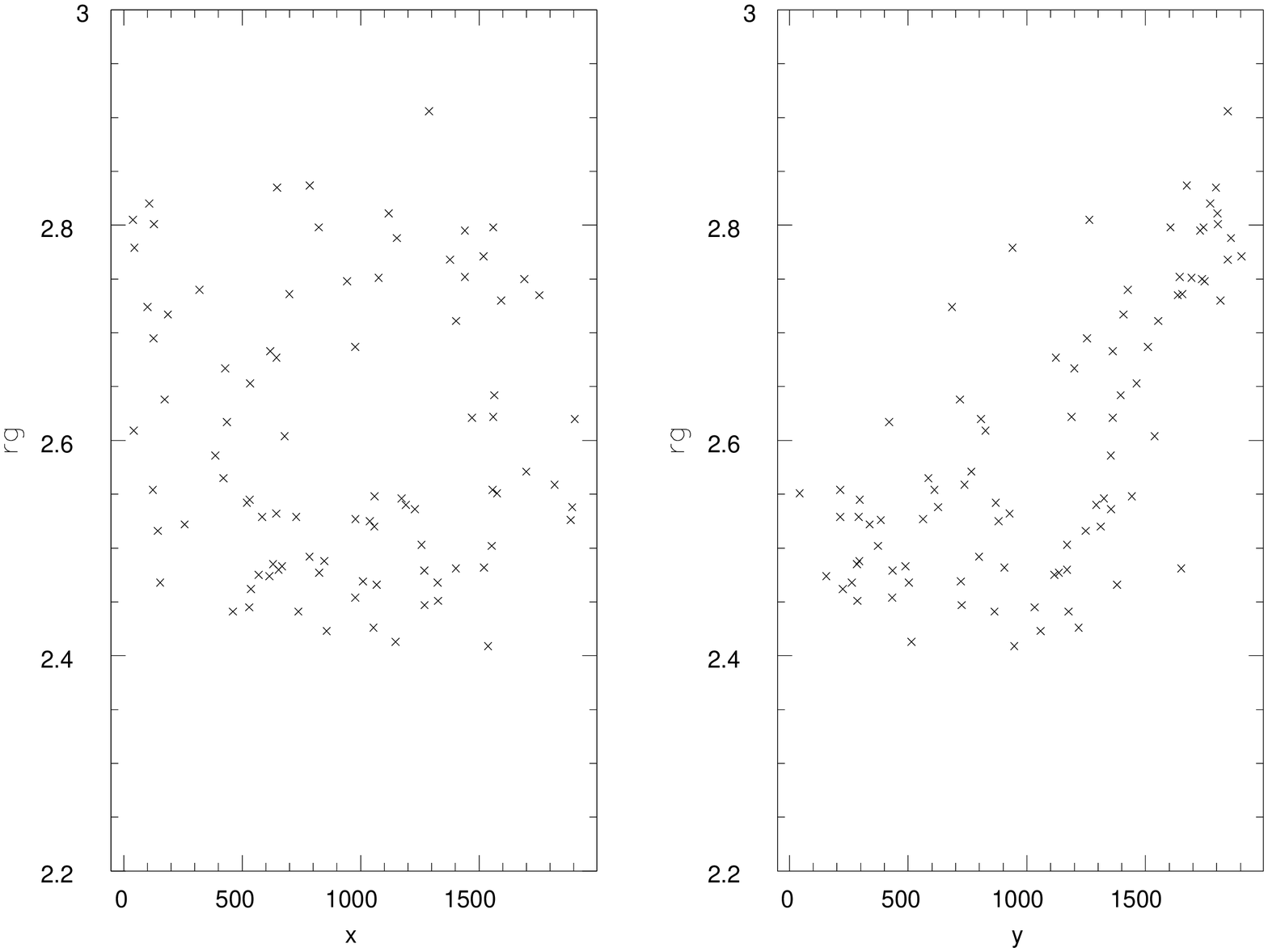}
\hfill
\parbox[b]{55mm}{
\caption{Variation of the size of the PSF along the x-axis (left
panel) and y-axis (right panel) of the CCD on EMMI. Note the increase
in size near the top edge of the CCD. The quantity $rg$ is the size of
the stellar images as determined by the KSB algorithm. The image from
which this result was derived has a seeing of $\sim 1$ arcsec.}
\label{PSFsize} }
\end{figure*}

Since there were reports (Erben 1996) of problems in the optics of EMMI
before its refurbishing, for comparison EMMI data from 1996 (Villumsen
\etal 1998) and test data taken in April 1997 have also been analyzed.
The PSF anisotropy map derived from images from 1996 showed 
erratic behavior even for consecutive exposures taken 10 minutes
apart. By contrast, EIS images using the refurbished EMMI have proven
to be quite stable. In fact, under similar observing conditions, the
PSF anisotropy shows no strong time-dependent variations, as can be
seen in Figure~\ref{stability} which displays nine consecutive frames
of an EIS OB. This stability implies that the strong optical
anisotropy can usually be corrected for.

\begin{figure*}[ht] \resizebox{12cm}{!}{\includegraphics{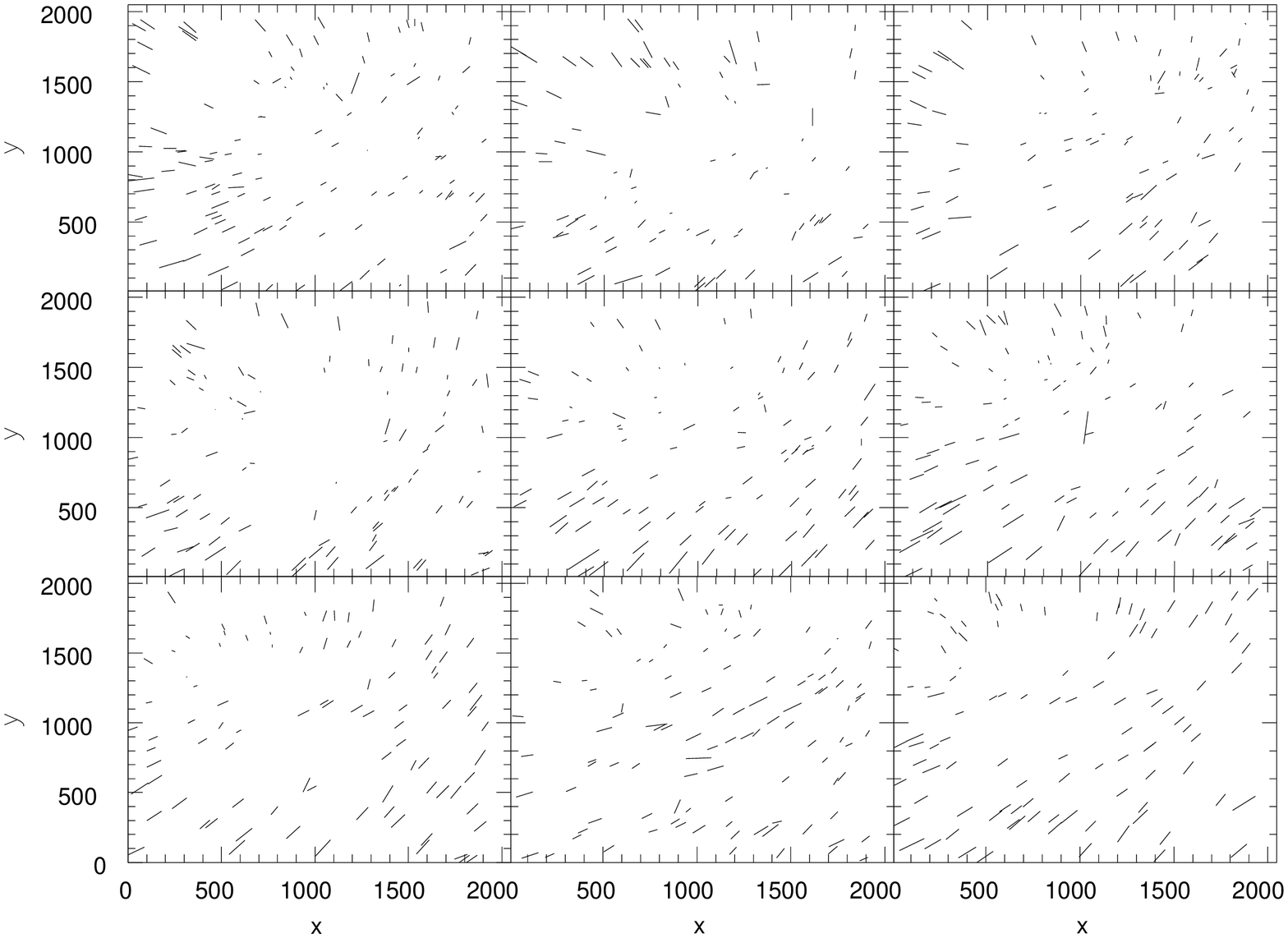}}
\hfill \parbox[b]{55mm}{
\caption{Nine consecutive EIS
images showing the stability of the PSF anisotropy with time.}
  \label{stability} }
\end{figure*}    

The continued monitoring of the EMMI PSF over an extended period of
time will provide valuable information in order to better understand
all the potential sources that may contribute to the anisotropy such
as the telescope tracking and pointing as well as environmental
effects. Even if this exercise proves to be of limited use for EIS,
the implementation of these tools may be of great value for future
surveys.

\subsection {EIS Database}
\label{database}

A survey project like EIS collects a large number of science and
calibration frames under varying conditions and produces a wealth of
intermediate calibration parameters, catalogs and images. This
multi-step process needs accurate monitoring as well as traceback
facilities to control the progress and steer the survey as a
whole. EIS is using a relational database consisting of several
tables, which have been implemented in the course of the ongoing
survey.

There are tables dedicated to storing parameters related to the
observations such as: 1) FITS keywords for all images delivered by the
ESO Archive to the EIS data reduction group; 2) Additional keywords
for different types of images which include survey frames (tiles),
photometric standards, astrometric reference frames, bias, flat-fields
and darks; 3) extinction data observed routinely by the Swiss
telescope and delivered by the Geneva Observatory.

Another set of tables are used to control and store the results of the
photometric calibration process including: 1) additional parameters of
the frames of photometric standards; 2) information and results from
different reduction runs of the photometric calibration data on a
frame by frame basis; 3) data for photometric and spectrophotometric
standards, combining results from the literature and from the
reduction of the calibration frames on a star by star basis.

Finally, there are tables that: 1) store basic information about the
nights in which EIS observations have been carried out; 2) control and
monitor the processing of the survey data; 3) store information about
the coadded sections and the catalogs produced by the pipeline.

All these tables are related by common keys and the underlying
commercial database engine used (Sybase) provides a powerful SQL
dialect to manage and retrieve the stored data.  Even though the
EIS-DB is far from complete, its implementation is important as one
can take full advantage of the available DB engine to retrieve key
information about the survey, to control the processing and to provide
a variety of statistics that can be used to fully characterize it. A
full description of the EIS database will be presented elsewhere (Deul
\etal 1998).

\subsection{Weighting and Flagging}
\label{weight}

\subsubsection{Rationale}

Because of the small number of EIS frames entering coaddition at a
given position in the sky (typically two frames), defects and other
undesired features cannot be rejected through robust combination such
as $\sigma-clipping$ or by taking the mode of a histogram. In
addition, two frames covering the same area in the sky can be observed
at different epochs with very different seeing. Therefore, artifacts
need to be identified on the individual images themselves, and
discarded from coaddition. This can be achieved by creating maps of
bad pixels.

This approach has been expanded to a more general one, which is the
handling, throughout the pipeline, of a set of weight-maps and
flag-maps associated with  each science frame. Weight-maps carry the
information on ``how useful'' a pixel is, while flag-maps tell why that
is so. Using this approach, the image processing task does not need to
interpret the flags and decide whether a given feature matters or not,
which improves the modularity of the pipeline.

\subsubsection{Implementation}

The creation of weight-maps and flag-maps is left to the WeightWatcher
program (see ``http://www.eso.org/eis/\-eis\_soft.html''). Several
images enter into the creation of the weight and flag-map associated
to each science frame. A gain-map --- which is essentially a flatfield
where the differences in electronic gains between the two read-out
ports have been compensated --- provides a basic weight-map. It is
multiplied by a hand-made binary mask where regions with very strong
vignetting (gain drop larger than 70\%) and a $\approx 30''\times
30''$ CCD coating defect are set to zero and flagged.  Bad columns
stand out clearly in bias frames. Affected pixels are detected with a
simple thresholding, marked accordingly in the flag-map, and again set
to zero in the weight-map. A thresholding suffices to identify
saturated pixels on science frames. These various steps are shown in
Figure~\ref{fig:mask}.

\begin{figure*}
\resizebox{8cm}{!}{\includegraphics{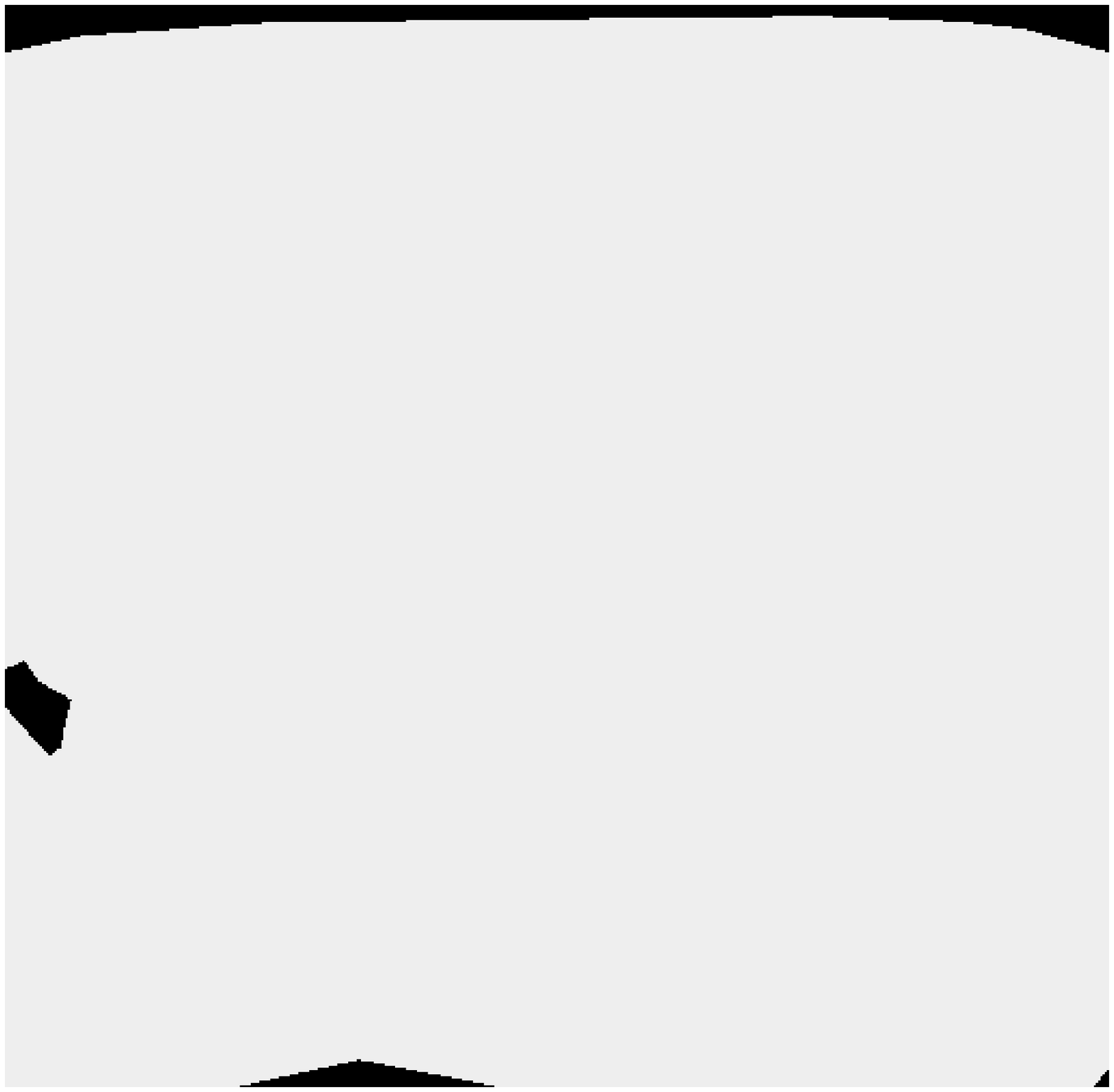}}
\resizebox{8cm}{!}{\includegraphics{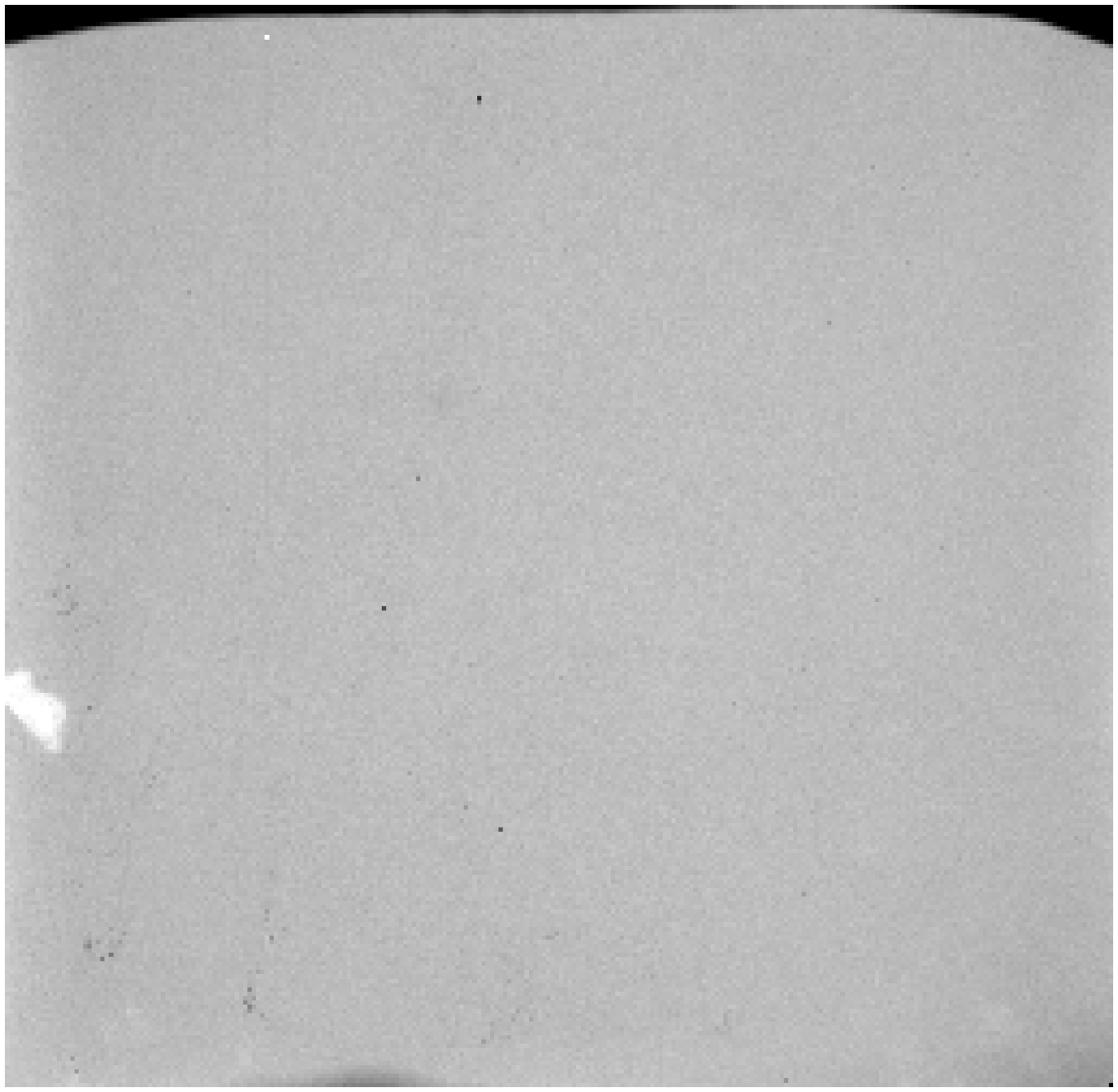}}
%\resizebox{8cm}{!}{\includegraphics{bias.ps}}
\caption[]{Two components of the  weight maps: the
hand-drawn mask (left panel), which excludes strongly vignetted
regions, and the gain map (right panel) obtained from the
flatfield. }
\label{fig:mask}
\end{figure*}

\subsubsection{Identification of electronic artifacts}

Glitches cannot be identified as easily. Those include cosmic ray
impacts, ``bad'' pixels and, occasionally, non-\-saturated features
induced by the intense saturation caused by very bright stars in the
field of view. Instead of designing a fine-tuned, classical algorithm
to do the job, a new technique has been applied based on neural
networks, a kind of ``artificial retina''. The details of this
``retina'' are described elsewhere (\cite{bertin:1998}), but it suffices
to say here that it acts as a non-linear filter whose characteristics
are set through machine-learning on a set of examples. The learning is
conducted on pairs of images: one is the input image, the other is a
``model'', which is  what one would like the input image to look
like after filtering.

\begin{figure*}
\resizebox{8cm}{!}{\includegraphics{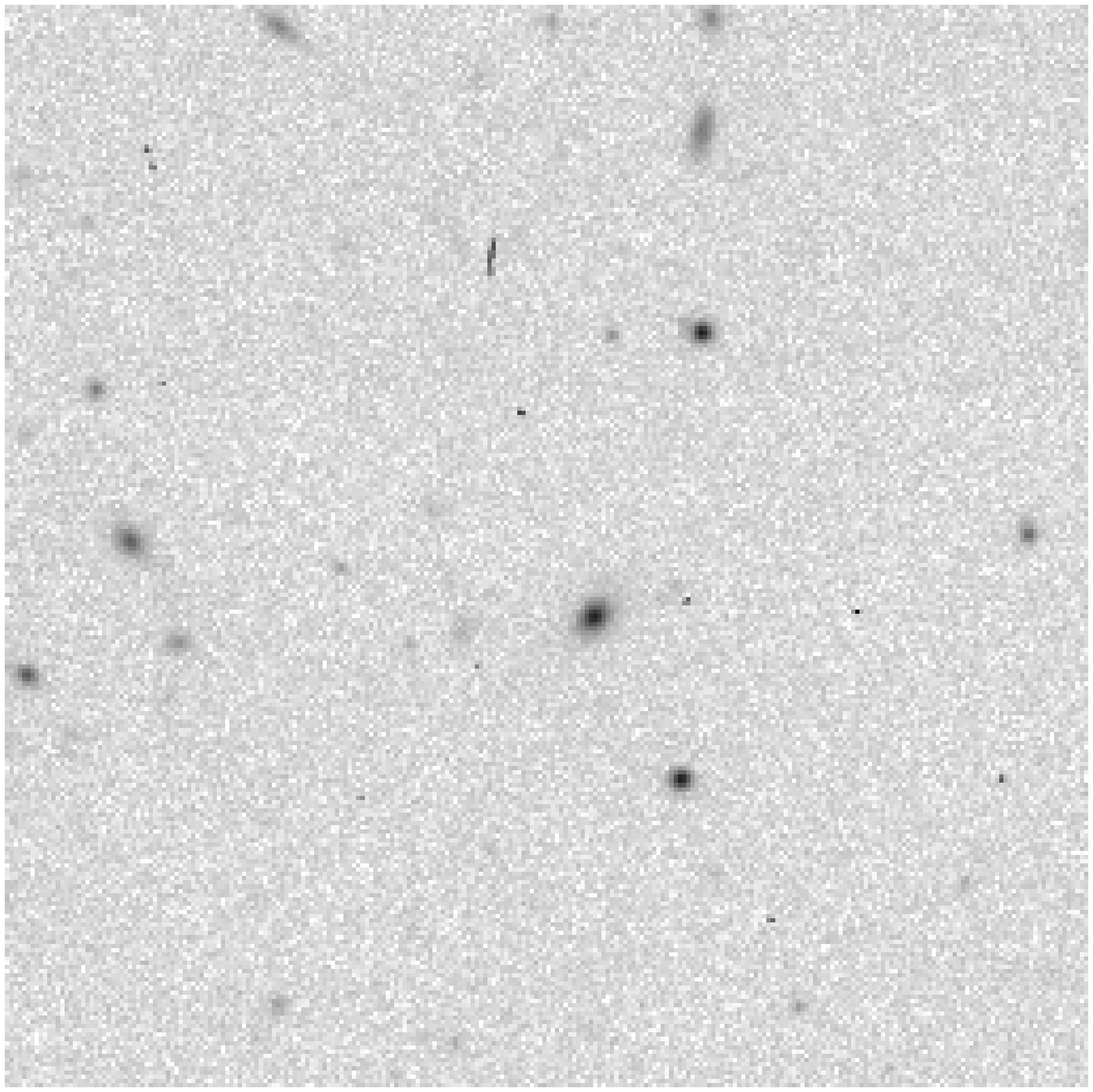}}
\resizebox{8cm}{!}{\includegraphics{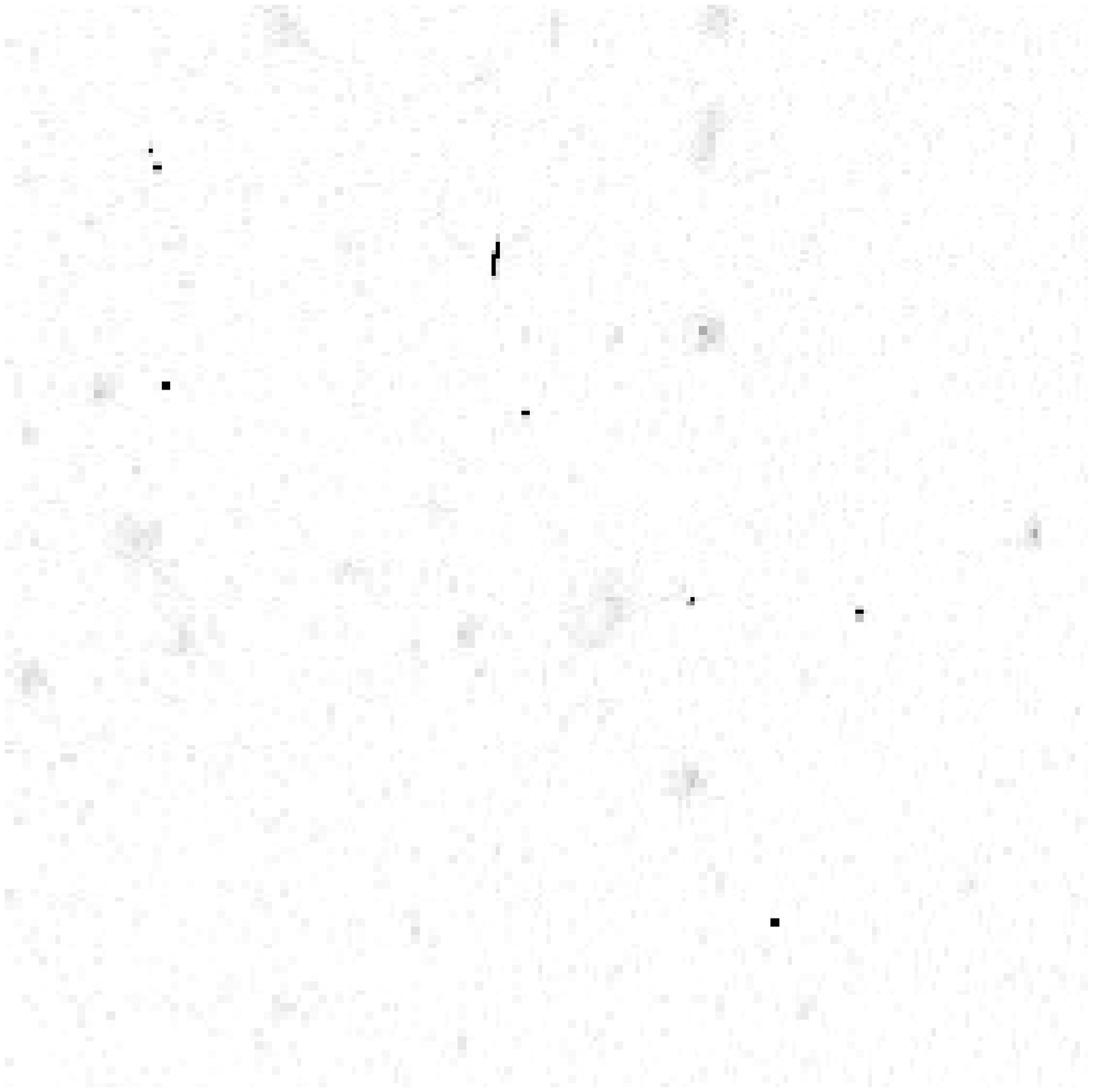}}
\caption[]{{\em Left}: Part of an EIS image with very good seeing
(FWHM$ = 0.54\arcsec$).{\em Right}: retina-filtered
image of the same field, showing up cosmic-ray impacts. Both images
with negative scale.}
\label{fig:retinex}
\end{figure*}

For the detection of glitches, a set of typical EIS images was used as
input images, reflecting different seeing and S/N conditions (but
putting stronger emphasis on good seeing images).  Dark exposures
containing almost nothing but read-out noise, cosmic-ray impacts and
bad pixels were compiled.  To these images a selection of more typical
features, induced by saturation, were added.  These ``artifact
images'' were then used as the model images, and were also added to
the simulations to produce the input images. The first EIS images used
as input already contain  unidentified cosmic-rays,
producing ambiguity in the learning. Thus the process was done
iteratively (3 iterations), using images where the remaining obvious
features are identified with a retina from a previous learning, and
discarded.  An example of a retina-filtered image is shown in
Figure~\ref{fig:retinex}.  

A crude estimate (through visual inspection) of the success rate in the
identification of pixels affected by glitches is $\sim $95\%. The
remaining 5\% originate primarily from the tails of cosmic-rays
impacts which are difficult to discriminate from underlying
objects. The spurious detections induced by these residuals are,
however, easily filtered out because of the large fraction of flagged
pixels they contain (see section \ref{single}).

\subsubsection{Other artifacts}

Unfortunately, there are other unwanted features that cannot be easily
identified as such automatically: optical ghosts and
satellite/asteroid trails. At this stage of the pipeline, obvious
defects of this kind were identified through systematic visual
inspection of all science images (section~\ref{frame}. The approximate limits of the
features are stored as a polygonal description which allows
WeightWatcher to flag the related pixels and put them to zero in the
weight-map.

\subsection{Astrometric Calibration}
\label{astrom}

To derive accurate world coordinates for the objects extracted from
EIS frames both the pairing of extracted objects in overlap regions
and pairing of extracted objects with a reference catalog (USNO-A1)
are used.

The pairing information of extracted objects with the reference
catalog is obtained by assuming that the image header information is
correctly describing (within 10\%) the pointing center and
the pixel scale of the image. Using a pattern recognition technique,
that allows for an unknown linear transformation, the pattern of
extracted objects is matched with the pattern of reference stars.
This results in corrections to the pointing center and pixel
scale. Then applying these corrections the pairing between extracted
objects in the overlap regions is performed.

\void{
To derive accurate world coordinates for the objects extracted from
EIS frames two types of information are used. The first is the overlap
regions between frames which provide the means to derive instrumental,
optical path deformations. Because astronomical objects are observed
many times, on different parts of the detector along different paths
through the optical system, deviations from a pure projection will
become evident.
For the second principle, we simultaneously map this
relative reference system to true world coordinates; this is done by
pairing the extracted objects with a position reference catalog
(USNO-A1). The latter mapping allows the derivation of global
deformation characteristics, while the first is better at deriving the
higher order terms of the plate deformation.}

Because the unit of measurement is a set of ten consecutive,
overlapping frames in which the telescope is operating in a
mechanically coherent manner (offset pointing, vs. preset pointing
between sets of ten frames) this forms the basis for an astrometric
solution. Independent pointing offsets are determined for each frame,
while considering the focal scale and distortion parameters as
constant or smoothly variable as a function of frame number in the
set. The plate model is defined as the mapping between pixel space and
normal coordinates (gnomonic projection about the field
center). This mapping is a polynomial description for which the
polynomial parameters are allowed to vary smoothly (Chebychev
polynomial) with frame number in the set.  Therefore, the mapping
between (x,y) pixel space and ($\zeta$, $\eta$) allows for flexure and
other mechanical deformation of the telescope while pointing.

Using pairing information, association of the source observed in several
frames, and association of the extracted source with the reference
catalog, a least squares solution is derived where these distances
between the members of the pairs are simultaneously minimized. Weighting
is done in accordance with the positional accuracy of the input data.
The source extraction accuracy rms is 0.03 arcsec, while the reference
has an rms of order 0.3 arcsec.  The astrometric least squares solution
is done in an iterative way. Because associations have to be derived
before any astrometric calibration has been done, erroneous pairings may
occur. A Kappa-Sigma clipping technique allows for discarding the large
distance excursions (probable erroneous pairings) between the
iterations.

From the astrometric solution it is found that distortion of the
pixel-scale is $\lsim 0.5$\% and the accuracy of the relative astrometry
is $\sim~0.03$ arcsec.

\subsection{Photometric Calibration}
\label{photom}

\subsubsection {Method}

Deriving a coherent photometric system for observations done in both
photometric and non-photometric conditions is a challenge handled in
EIS through a stepwise procedure.

The first step is to derive the relative photometry among frames in an
overlapping set. In contrast to what is done for the astrometric
calibration, all overlaps among frames in a contiguous sky area are
used. Using the overlap pairing information, an estimate for the total
extinction difference between frames can be computed. This can be
expressed in terms of a relative zero-point, which by definition,
includes the effects of airmass and extinction. To limit the magnitude
difference calculation to reliable measurements, a selection of input
pairs is made based on the SNR, maximum allowed magnitude difference
between members of a pair, and limiting magnitudes for the brightest
and faintest usable pairs. Because a set of frames will have multiple
overlaps, the number of data points (frame-to-frame magnitude
differences) will be over-determined with respect to the number of
frames. Therefore, the relative zero-point for each frame can be derived
simultaneously in a least squares sense. Weighting is applied based on
the number of extracted objects and their fluxes.
The solution is computed in an iterative fashion. Estimated
frame-based zero-points from a previous iteration are applied to the
magnitudes and new sets of pairs are selected, rejecting extraneous
pairs. The process stops when no new rejections are made between
iterations.  The internal accuracy of the derived photometric solution
is $\lsim 0.005$ mag.

The second step involves correcting possible systematic photometric
errors, and deriving an absolute zero-point. Systematic errors are
introduced by incorrect flat-fielding (stray-light, pixel-scale
variations, gain changes between read-out ports) or variation of image
quality. The latter has, however, been minimized by adopting an
appropriate photometric estimator (section~\ref{measure}). The
correction for these systematic errors can be made by using external
information, such as pointed measurements from other telescopes,
and/or absolute zero-points from EIS measurements of Landolt
standards, which are used to anchor a global photometric solution. As
long as these observations cover the patch uniformly, systematic
zero-point errors can be corrected by a weighted least-square fit of a
low-order polynomial to the difference between the relative zero-point
derived from the previous step and the external pointed measurements.
This general procedure will be adopted in the final version of the EIS
data.

\subsubsection {Calibration of patch A}

 In this preliminary release, the zero-point calibration of patch A
was determined to be simply an offset derived from a weighted average
of the zero-points of the available anchors.  For patch A, the
available anchor candidates are: the 2.2m data, the 0.9m data and the
EIS tiles taken during photometric nights. Given the problems detected
with the 2.2m data, the latter have been, for the time being,
discarded. For patch A, five nights were observed under photometric
conditions. This represents about one hundred tiles, covering a wide
range in declination (Figure~\ref{fig:overlaps}).

\void{Unfortunately, the external photometric data that can be used to
anchor the absolute zero-point of patch A are not evenly
distributed in the surveyed region, as shown in Figure~\ref{fig:overlaps}. 
This is a consequence of the unexpectedly bad weather during the
visibility period of patch A.}

There are some indications for a small zero-point gradient in right
ascension ($\sim$ 0.02~mag/deg), in agreement with the uncertainties
of the flat-fields ($\sim 0.002$ mag). By contrast, the behaviour
along the north-south direction is much less well determined, as it
relies on calibrations generally carried out in different nights. With
the current calibrations, however, the systematic trend is estimated
to be $\lsim$~0.2~mag peak-to-peak.

The DENIS strip could have been a perfect data set to constrain the
homogeneity of the zero-point in declination. However, careful
examination of the the DENIS standards revealed a significant
variation in the zero-points for the night when this strip was
observed, preventing its use to constrain possible gradients in
declination of the EIS data. Note that another strip crossing the
surveyed area has also been observed and an attempt will be made to
retrieve it from the DENIS consortium before the final release of the
EIS data.

\void{
As a final check of the EIS magnitude system, the EIS catalogs
extracted from the calibrated data are compared to the catalogs
produced by the DENIS survey and the catalog generated using the same
software as in the EIS pipeline from the images available from Lidman
\& Peterson (1996).  The magnitude difference for the objects in
common in the EIS and DENIS catalog is shown in
Figure~\ref{fig:denis}, as a function of the EIS magnitude.  The
comparison yields a mean offset of 0.17 mag with an rms of about 0.05
mag (for EIS magnitudes $12 < {\rm I} < 16$ and excluding the obvious
erroneous pairings and saturated EIS data points at I$<15$), taken
hereafter as our nominal error in the zero-point of the patch A
absolute scale. For EIS magnitudes I $<17$ and excluding the obvious
erroneous pairings and saturated EIS data points at I$<15$, the values
become 0.17 mag for the offset and 0.11 for the rms.  The rms of the
fit is dominated by the errors of the DENIS photometry as the DENIS I
limiting magnitude is $18.5$ mag and EIS objects brighter than
$15^{\rm th}$ mag are over-exposed.}

\void{
 The
distribution implies that while one may be able to impose constraints
along the declination, possible systematic variations along right
ascension are largely unconstrained. Note that there is another,
unprocessed DENIS strip crossing patch A at a different location.
Using those data it
will be possible to further check the absolute photometry before the
final release of the EIS data.}

\void{
Because of the difference in sensitivity between the EIS and DENIS
surveys, there is only a small magnitude range over which we can check
the absolute photometry ($16 < I < 18$). The one DENIS strip,
currently reduced, that crosses the EIS patch A has been used for the
comparison in Figure~\ref{fig:denis}. This comparison was made after
removing the linear systematic variations of the zero-point. The
scatter in Figure~\ref{fig:denis} is largely due to the photometric
errors in the DENIS data and the difference between the two
photometric systems.}

\subsection {Coaddition}

Coaddition serves three main purposes: it increases the depth of the
final images; it allows the suppression of artifacts which are present
on some images (such as cosmic-ray hits); and it allows the creation of
a "super-image" of arbitrary size to facilitate studies which would be
affected by the presence of the edges of the individual frames.

A general coaddition program had been developed from the "drizzle" method
originally created to handle HST images (\cite{fruchter}, \cite{hook}). 
The inputs for each coaddition operation are the data frame
and a weight map which also includes information about pixels which have
been flagged as invalid. The corner positions of each input pixel are
transformed, using the astrometric solution generated by the pipeline, first
into equatorial sky positions and then onto pixel positions in the output
super-image. The conic equal-area sky projection is used for the super-image
as it minimizes distortion of object shapes and areas. To ease manipulation
and display the super-images are normally stored as sets of contiguous
4096x4096 pixel sections.

The data value from the input is averaged with the current values in the
output super-image using weighting which is derived by combining the weight
of the input, the weight of the current output pixel and the overlap area
of the projection of the input pixel onto the output grid. The method
reconstructs a map of the surface intensity of the sky as well as an output
weight map which gives a measure of the statistical significance of each
pixel value. A third output image, the "context map" encodes which of the
inputs were coadded at a given pixel of the output so that properties of
the super-image at that point, such as PSF, can be reconstructed during
subsequent analysis. The values in the context-map provide pointers to a
list which is dynamically generated and updated during coaddition. This is a
table of the unique image identifiers for a specified context as well as the 
number of input images and the number of pixels having a given context value.

\section {Data Products}

The EIS pipeline produces a wide array of intermediate products, both
images and catalogs, as well as a large set of logs, reports and
diagnostics.  A full description of these products is beyond the scope
of the present paper and will be presented elsewhere.

In this preliminary release the following data are publicly available
at ``http://www.eso.org/eis/datarel.html'':

{\bf Single Frames:} sky-subtracted, fully calibrated frames, in
integer format. They can be retrieved at the above WWW address, where
a request form will be available. The requests will be handled by the
Science Archive Group. Note that the calibrated images being
distributed have not been corrected for cosmetic defects. In
particular, they contain bad columns, cosmic rays and the vignetted
region at the top and bottom edges of the frames.  However, all of
these artifacts have been taken into account in the production of the
object catalogs and the information available in them allows one to
filter out affected objects.  To provide approximate astrometric
information the CD-matrix convention (as given in the Users Guide for
the Flexible Image Transport System 1997 version 4) has been adopted
to describe the world-coordinate system, and included in the FITS
header of each frame. Even though this is in general a reasonable
approximation it is not appropriate to account for the distortions of
the frames. Therefore,when object catalogs are overlayed onto the
images some residuals are visible, especially at the corners of the
images. The photometric zero-point for each frame after the absolute
photometric calibration of the frame appears in the header of the
image. A full description of the EIS specific keywords can be found at
``http://www.eso.org/eis/keywords''.

{\bf Coadded Sections:} sky-subtracted, fully processed (see above)
sections of the coadded image. These sections are mapped using the
conic equal area (COE) projection (Greisen \& Calabretta 1996). Note
that this projection is not handled by all display-tools but has
already been implemented in the ESO SkyCat.  The available images are
lossy compressed using a HCOMPRESS library originally written by White
(1992) for the STScI. The coadded sections have been compressed using
a very high compression scale of 200. In addition to the compression,
the images are re-sampled by a factor of 0.5 in both directions using
the re-sampling code developed by Devillard (1997). The size of the
original images is 67 MB, whereas that of the compressed and
re-sampled images is about 150 kB. These sections are handled by an
on-line server, similar to that available for the Digitized Sky Survey
(DSS).  A preliminary object catalog, containing only position and
shape information is also available on the on-line server. For the
display of the sections and the catalog SkyCat can be used as an
interface.

{\bf Low-resolution Coadded Image:} sky-subtracted, fully processed
coadded image. This image is meant to give a general overview of the
whole patch and it has been produced by re-binning (3 arcsec pixel
size) and pasting together the sections.  This is a standard output of
the pipeline. This image ($\sim 25$Mb) can be retrieved at the same
web page as above.

{\bf Single Frame Catalogs:} Object catalogs associated with each
single frame. A full description of the parameters available is
presented below. The name convention of the catalogs is based on that
of the EIS tiles $A_{i,j}$ (see section 2). The catalogs are in binary
FITS table format.

\void{
In addition to the products described above, an example of a single
frame image at full resolution and the corresponding filtered catalog
(see below) are also available at the same web location.}

Note that the coadded sections and image currently available are not
suited for astronomical reduction because of the varying noise
properties and compression, but are useful for proposal preparation
and comparison with EIS catalogs or other data sets.

The main difference between the current release and the final release of
the EIS data is that the latter will also include the coadded image and
the auxiliary weight and context maps, as well as the final object
catalog which will contain information about the context within which a
given object has been identified. The inclusion of the context
information into the catalog extracted from the coadded image is
currently underway.

\section {Object Catalogs}

\subsection{Source Detection and Photometry}

In a survey like the EIS, where a large variety of astronomical ---
and non-astronomical!--- objects of all kinds can be detected and
measured over wide areas, one cannot avoid making choices.  In the
case of EIS, the priority is the detection of objects such as faint stars
and galaxies. Brighter objects are generally saturated and/or already
cataloged.  The source extraction is performed with a new version of
the SExtractor software (\cite{bertin}) that can be retrieved from
``http:/www/eso.org/eis/eis\_soft.html''. SExtractor is optimized for
large scale imaging survey fields with low to moderate source density,
and is therefore perfectly suited to EIS. The processing is done in 3
steps: detection, measurement, and classification, which are briefly
described for the single image process.

\subsubsection {Detection}

For each image, the detection process in SExtractor begins with the
determination of a smooth background map. This is done by computing
the modes of histograms built from meshes of $64 \times 64$ pixels,
corresponding to $\approx 17$ arcsec. This relatively small scale was
chosen in order to facilitate the detection of faint objects on top of
the strong gradients encountered near the many bright stars in the
survey (3 out of the 4 EIS-wide fields are located at moderate
galactic latitude). This produces a lower resolution image ($32 \times
32$) of the background (hereafter referred to as miniback).  This
miniback is median-filtered using a $3 \times 3$ box-car, to avoid the
contamination of the background map by isolated, extended objects.
The median filtering also helps to reduce photometric bias for
bright, ``large'' galaxies, to a negligible fraction up to scales $\ga
40$\arcsec, which correspond to $I \la 18$. A full-resolution
background map is obtained by interpolating the smoothed miniback
pixels, using a bicubic-spline, and is subtracted from the science
image.

Background-subtracted images are then filtered before being
thresholded, to reduce the contribution of noise on spatial scales of
the image where it is dominant. The median seeing (FWHM) of EIS
images as a whole is about 0.9 arcsec, a little more than 3 pixels. The
data are filtered by convolving with a slightly larger, constant,
Gaussian profile with FWHM~$= 4$~pixels. Although the choice of a
convolution kernel with constant FWHM may not always be optimum (the
seeing may vary by as much as a factor of 3), the impact on
detectability is, however, fairly small (see \cite{irwin}). On the
other hand, it has the advantage of requiring no change of the
relative detection threshold. It also simplifies the comparison with
the coadded-image catalog, for which the convolution kernel is also
fixed.

The detection threshold, $k\sigma$, used in SExtractor is expressed in
units of the standard deviation $\sigma$ of the background noise. For
single images $k = 0.6$ is used, which corresponds to a typical
limiting surface brightness $\mu_I \sim 24-24.5$.  The new SExtractor
allows this noise-level to be set independently for each pixel $i$,
using a weight-map $w_i$ (Section~\ref{weight}), which is internally
converted to a relative variance: $\sigma^2_i \propto w^{-1}_i$.  The
variable detection threshold is also used for deciding if a faint
detection lying close to a bright object is likely to be spurious or
not.

Some pixels are assigned a null weight by WeightWatcher, because they
are unreliable: gain too low, charge bleeding, cosmic-ray, etc. The
detection routine cannot simply ignore such pixels, because some
objects, like those falling on bad columns or charge bleeding
features, would be either truncated or split into two.  A crude
interpolation of bad pixels overcomes this problem. Unfortunately,
interpolation creates correlated patterns which are sometimes detected
at the very low thresholds applied in the EIS, but as these zones are
flagged, they are easily filtered out in the final catalog.

\subsubsection {Low Surface Brightness Features}

The above procedure is primarily sensitive to objects with sizes
smaller or comparable to the background mesh size. In order to
properly detect objects on larger scales, the minibacks are used
instead of the original images. By using the minibacks the scale of
the objects that can be detected can be extended to a significant
fraction of a frame. The largest scale for detection is imposed by the
flatfield inhomogeneity and by extended halos around bright stars.

These minibacks are being examined systematically to produce a catalog
of low surface brightness objects which will be released at a later
date (Slijkhuis \etal 1998). For current EIS frames the scales probed
range from $\sim 17$ arcsec to roughly 140 arcsec.

\subsubsection {Measurement}
\label{measure}

Basic positional and shape parameters are computed for each detection
on the convolved image. These include the baricenter, major, minor
axes and position angle derived from the second-order moments of the
light distribution and the associated error-estimates, which take into
account the weighting of each pixel. The photometry is performed on
the un-convolved, un-interpolated image. Photometric parameters measured
on the images include isophotal magnitudes, fixed-disk aperture
magnitudes with diameters ranging from 2.7 to 14 arcsec (14 arcsec is
the typical ``Landolt aperture''), and SExtractor's estimate of
``total'' magnitude: MAG\_AUTO. The latter is a Kron-like elliptical
aperture magnitude. It is computed in a way similar to that proposed
by Kron (1980), except that the aperture is required to be elliptical,
with aspect-ratio and position angles derived from the second-order
moments. For the measurement of magnitudes, pixels with zero-weight
and those associated with the isophotal domain of some neighbor are
handled in a special way by the new SExtractor: when possible, they
are replaced by the value of the pixel symmetrical to the current one,
with respect to the baricenter of the object.  Although this simple
algorithm is certainly crude, it proves to yield fairly robust results
and replaces advantageously the MAG$\_$BEST estimator used in the old
SExtractor (Bertin \& Arnouts 1996). One particular aspect of EIS is
the large variation in the seeing from frame to frame.  Simulating
EIS images of points sources under different observing conditions, it
is found that the MAG$\_$AUTO magnitudes are fairly robust with respect
to seeing variations: systematics of only $\approx 1$\% peak-to-peak
are expected for the bright stars used in the photometric solution.

\subsubsection {Classification}

The standard SExtractor star/galaxy classifier is a multilayered
back-propagation neural-network fed with isophotal areas and the peak
intensity of the profile. The classifier was trained with simulated
ground-based, seeing-dominated, optical images. It will therefore
perform well on images close to the conditions met in the original
simulations. This is so for EIS images in patch A, but it is no longer
the case for other patches, where very good seeing and strong optical
distortions yield significantly elongated and skewed stellar profiles,
varying over the frame. A new, more general, star/galaxy separation
scheme is therefore needed for these fields, and is currently being
implemented in SExtractor.

The current classifier returns a ``stellarity index'' 
between 0 and 1. A value close to 0 means the object is extended
(galaxy), while a value close to 1 indicates a point-source (star). It
can  be shown that the neural network output is approximately the
probability that an object is a point-source. This
is only valid for a sample of profiles which would be drawn from the
same parent population as the training set. Because the neural
classifier is a finely tuned system, these conditions are almost never
met with real images, and care has to be taken when interpreting the
stellarity index. Nevertheless, it is fair to adopt a stellarity index
value of 0.5 as a default limit between point-sources and elongated
objects. At faint levels (I$\gsim 21$), star/galaxy separation begins
to break down for frames obtained under the least favorable seeing
conditions in patch A. A clump begins to form around a stellarity
index of 0.5, indicating that the algorithm cannot provide a reliable
classification for most objects.

In the discussion below two values of the stellarity index are adopted
to separate stars and galaxies: the conservative value of 0.5, which
tends to favor more complete star catalogs, and a value of 0.75, which
assumes that beyond the classification limit galaxies largely
outnumber stars.

\subsection {Single Frame Catalogs}
\label {single}

The most basic catalogs are the single frame catalogs which are
generated by SExtractor. These are produced by default by the pipeline
in a two-step process. First, SExtractor is run with a high threshold
to identify stars and determine a characteristic value of the FWHM for
each frame. This value of the FWHM is then used as input to a second
run of SExtractor with a low-threshold for detection which also
provides the classification of the detected objects by computing the
stellarity-index.

During the extraction SExtractor sets several flags to describe any
anomalies encountered.  The meanings of the flags are summarized in
table ~\ref{flag_table}. Information available in the flag-maps
generated by the WeightWatcher program are also propagated to the
catalog.  Flags are set to indicate that a given object is affected by
bad pixels in the CCD-chip or by artifacts in the image that have been
marked either by the artificial retina or by the polygon-masking
during visual inspection (section~\ref{eyeinspection}).  Table
~\ref{imaflags_table} summarizes the meaning of the flags in the
catalog set from the information contained in the flag-maps.

\begin{table}[h]
\caption{Description of SExtractor Flags ($f_s$)}
\label{flag_table}
\begin{tabular}{rp{7cm}}\hline
Value & Description \\ \hline
1 & The object has neighbors bright and close enough to significantly
bias the MAG\_AUTO photometry\\
2 & The object was originally blended with another one \\
4 & At least one pixel of the object is saturated \\
8 & The object is truncated (too close an image boundary) \\
16 & Object's aperture data are incomplete or corrupted \\
32 & Object's isophotal data are incomplete or corrupted \\
64 & A memory overflow occured during de-blending \\
128 & A memory overflow occured during extraction \\
\hline
\end{tabular}
\end{table}

\begin{table}[h]
\caption{Description of WeightWatcher Flags ($f_w$).}
\label{imaflags_table}
\begin{tabular}{rp{7cm}}
\hline
Value & Description \\ \hline
1 & The object contains pixels that were marked in the image mask \\
2 & The object contains pixels with a deviating gain \\
4 & The object contains pixels with a deviating bias \\
8 & The artificial retina detected a cosmic ray hit within the object \\
16 & The object contains saturated pixels \\
\hline
\end{tabular}
\end{table}

The contents of the catalogs include: J2000.0 right ascension and
declination, x and y coordinates in the chip; total magnitude
(MAG\_AUTO) and error; major and minor axis; position angle;
stellarity index; SExtractor flag $f_s$ (see table \ref{flag_table});
WeightWatcher flag $f_w$ (see table \ref{imaflags_table}); total
number of pixels above the analysis threshold (npix); total number of
pixels that are flagged by WeightWatcher (nflag). Further information
can be found at ``http://\-www.eso.org/\-eis/eis\_soft.html''.

\void{
\begin{table*}
\caption{Bright Object Catalog}
 \begin{tabular}{llllllllllllll}
\hline \hline
\multicolumn{14}{l}{\#   1 ALPHA\_J2000     Derived Right Ascension J2000                  [deg]}\\
\multicolumn{14}{l}{\#   2 DELTA\_J2000     Derived Declination J2000                       [deg]}\\
\multicolumn{14}{l}{\#   3 XImage          Object position along x                         [pixel]}\\
\multicolumn{14}{l}{\#   4 YImage          Object position along y                         [pixel]}\\
\multicolumn{14}{l}{\#   5 Mag             Kron-like elliptical aperture magnitude         [mag]} \\ 
\multicolumn{14}{l}{\#   6 MagErr          RMS error for AUTO magnitude                    [mag]} \\
\multicolumn{14}{l}{\#   7 A               Major axis object ellipse in WCS                [deg]}\\
\multicolumn{14}{l}{\#   8 B               Minor axis object ellipse in WCS                [deg]}\\
\multicolumn{14}{l}{\#   9 Theta           Position angle object ellipse in WCS            [deg]}\\
\multicolumn{14}{l}{\#  10 Class\_Star      S/G classifier output}\\                          
\multicolumn{14}{l}{\#  11 Flag            Extraction flags}\\
\multicolumn{14}{l}{\#  12 Imaflags\_iso    FLAG-image flags OR'ed over the iso. profile}\\
\multicolumn{14}{l}{\#  13 Npix            Isophotal area above threshold                   [pixel$**$2]}\\
\multicolumn{14}{l}{\#  14 Nimaflags\_iso   Number of flagged pixels entering IMAFLAGS\_ISO}\\
\\
\tiny
341.490473 & \tiny -39.498289 & \tiny 1769.220  & \tiny 99.601 & \tiny 14.3569 & \tiny 0.0010 & \tiny 0.000251001 & \tiny 0.000220953 & \tiny -5.96764 & \tiny 0.98 & \tiny 0 & \tiny 32  & \tiny 926 & \tiny 0 \\
\tiny 341.363135 & \tiny  -39.494171 & \tiny  440.529 & \tiny  37.315 & \tiny  18.2139 & \tiny  0.0105 & \tiny  0.000170945 & \tiny  0.00016276 & \tiny  -43.1517 & \tiny  0.98 & \tiny  0 & \tiny  0 & \tiny  112 & \tiny  0 \\
\tiny 341.411834 & \tiny  -39.493039 & \tiny  949.491 & \tiny  24.493 & \tiny  22.8966 & \tiny  0.3517 & \tiny  3.70163e-05 & \tiny  3.68999e-05 & \tiny  51.4425 & \tiny  0.37 & \tiny  16 & \tiny  0 & \tiny  2 & \tiny  0 \\
\tiny 341.340957 & \tiny  -39.493305 & \tiny  208.497 & \tiny  24.491 & \tiny  23.1433 & \tiny  0.3483 & \tiny  3.70635e-05 & \tiny  3.67203e-05 & \tiny  50.2485 & \tiny  0.61 & \tiny  16 & \tiny  8 & \tiny  3 & \tiny  1 \\
\tiny 341.516308 & \tiny  -39.492140 & \tiny  2038.520 & \tiny  18.000 & \tiny  22.5124 & \tiny  0.2256 & \tiny  0.000170742 & \tiny  2.1373e-05 & \tiny  0.246188 & \tiny  0.05 & \tiny  16 & \tiny  3 & \tiny  8 & \tiny  8 \\
\tiny 341.417734 & \tiny  -39.492020 & \tiny  1011.136 & \tiny  11.027 & \tiny  20.3983 & \tiny  0.0772 & \tiny  0.000206869 & \tiny  0.000170264 & \tiny  39.1945 & \tiny  0.01 & \tiny  16 & \tiny  0 & \tiny  67 & \tiny  0 \\
\hline \hline
\end{tabular}
\label{tab:cats}
\end{table*}
}

\subsection {Derived Catalogs}

\label{sectderived}

During the processing of a patch through the pipeline the single frame
catalogs are merged together into a ``patch'' catalog which contains
information of all objects identified in the individual frames. Note
that objects may have multiple entries if they are in overlapping
frames. From this patch catalog several single entry catalogs may be
derived, for instance, the even/odd catalogs containing all objects
detected in the even/odd frames.  Objects detected in more than one
frame are identified to produce a single-entry in the final catalog,
choosing the parameters as determined from the best-seeing
image. Objects in regions of overlap are paired whenever the
baricenter of the smallest falls within an ellipse twice the size of
the object ellipse of the larger one. Details on this procedure will
be presented elsewhere (Deul \etal 1998).

From the flag information available in the single-entry catalog, {\it
filtered} catalogs can be produced for analysis purposes (see
Section~\ref{results}).  The filtering is required to eliminate
truncated objects and objects with a significant number of pixels
affected by cosmics and/or other artifacts. Objects with the following
characteristics are discarded: $f_s \geq 8$ or nflag/npix $ \ge 0.1$,
where $f_s$ is the SExtractor flag, npix the number of pixels above
the analysis threshold and nflag the number of pixels flagged by
WeightWatcher.  The two-dimensional distribution of stars and galaxies
from the resulting catalog are shown in Figures~\ref{fig:diststar} and
\ref{fig:distgal}, for different limiting magnitudes.

\begin{figure*}
\resizebox{15cm}{!}{\includegraphics{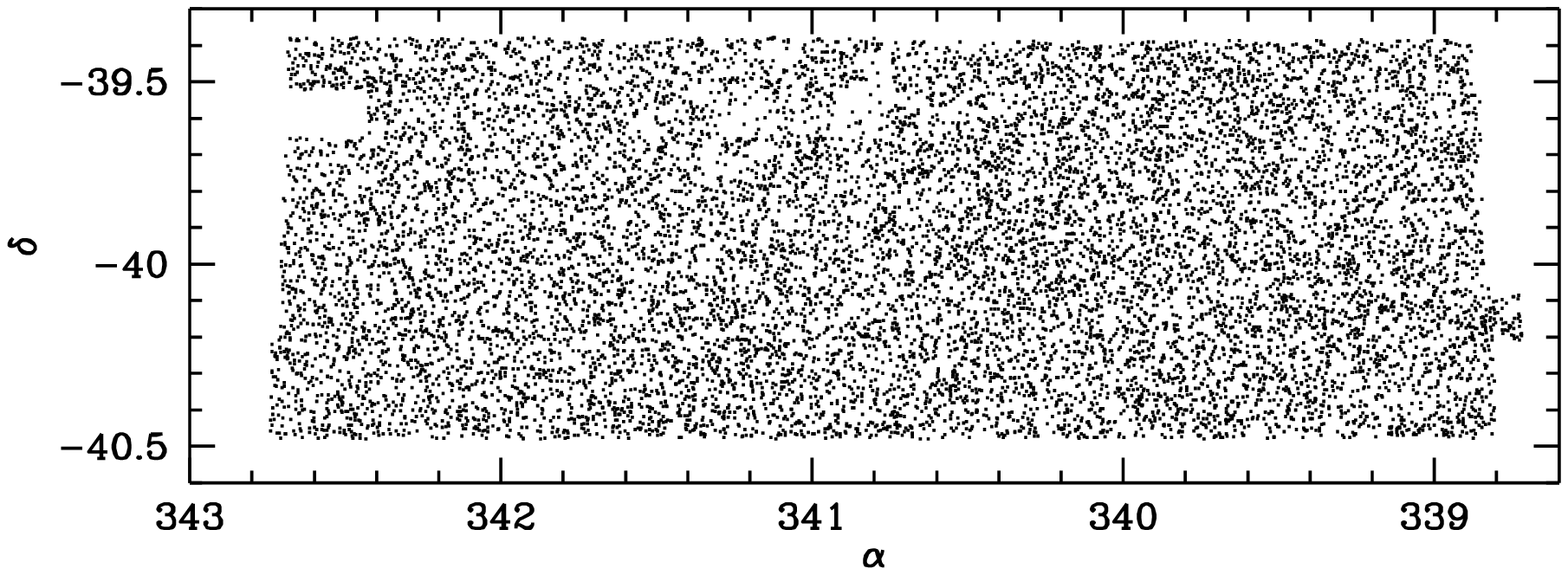}}
\resizebox{15cm}{!}{\includegraphics{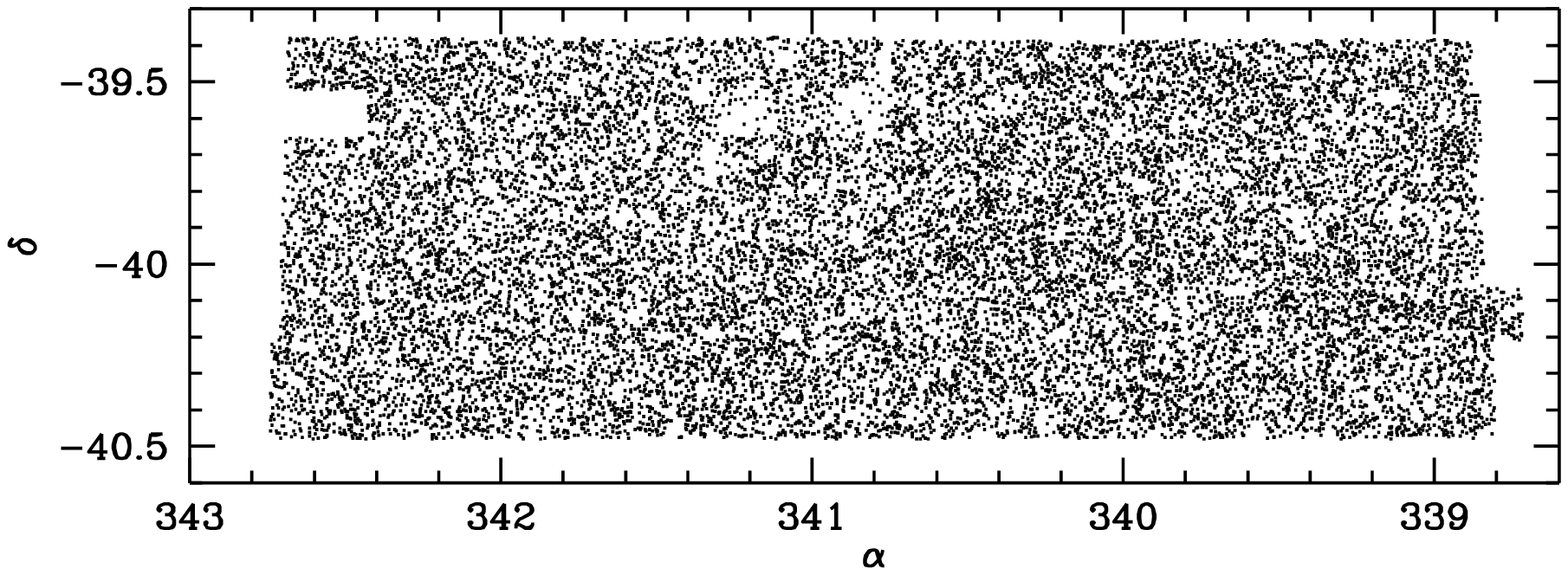}}
\caption[]{Distribution of stars detected in the even frames covering 
 patch A at magnitudes brighter than I$=20$ (12355 objects, upper
panel) and I$=21$ (18529, lower panel). Objects with a stellarity
index $>0.5$ were classified as stars. Note the two bad frames in the
upper part of the patch, yielding nearly empty regions. As pointed out
in the text this region is usually discarded from the analysis.}
\label{fig:diststar}
\end{figure*}

\begin{figure*}
\resizebox{15cm}{!}{\includegraphics{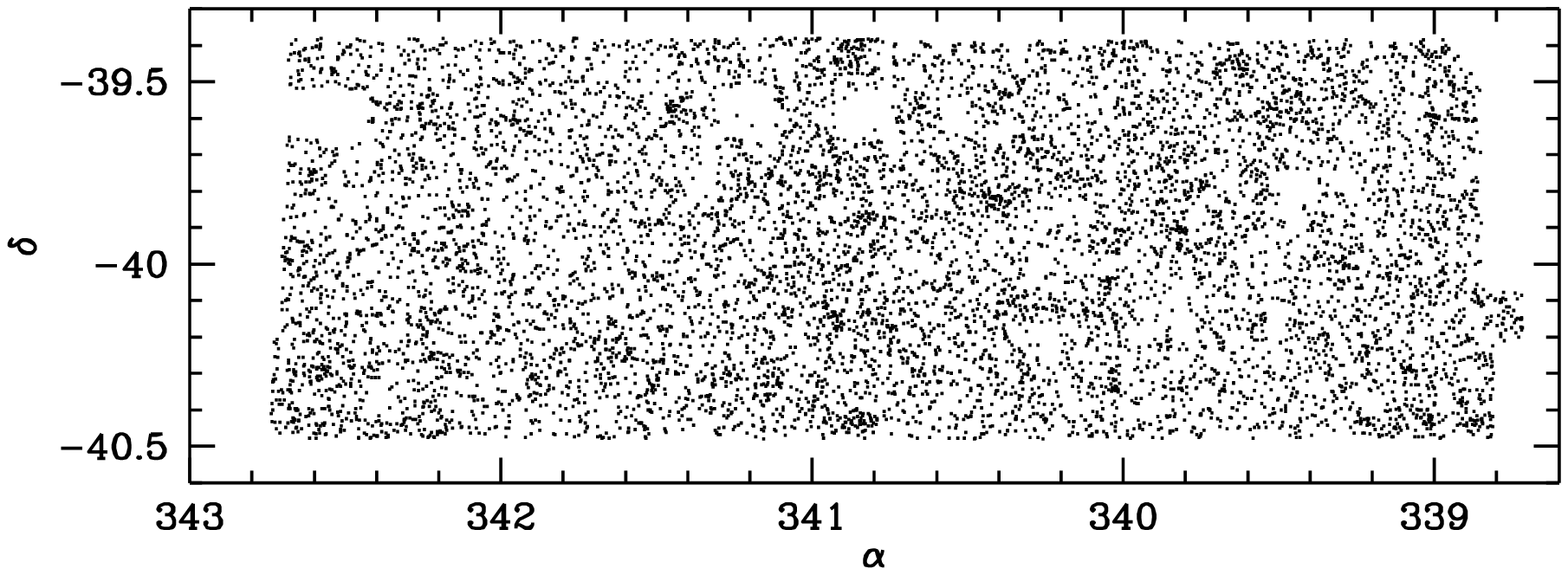}}
\resizebox{15cm}{!}{\includegraphics{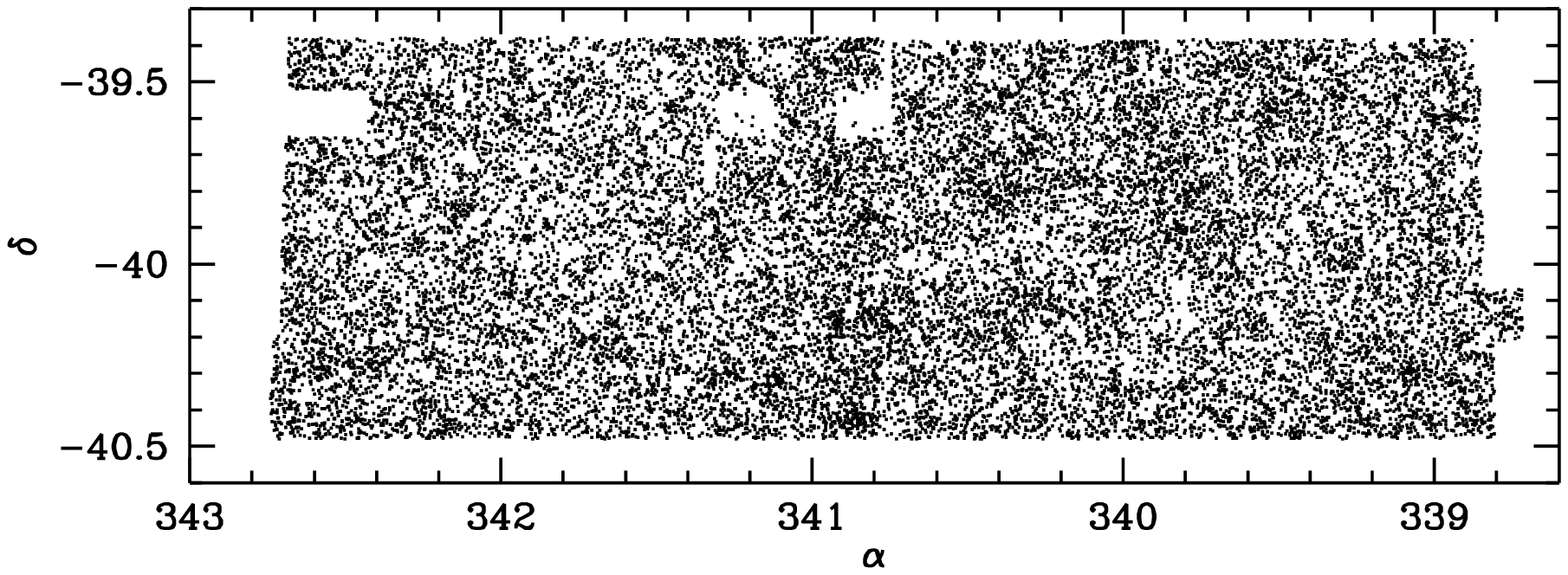}}
\caption[]{Same as in previous figure showing the distribution of
9006 (upper panel) and 23129 (lower panel) galaxies for the same two
limiting magnitudes as in Figure~\ref{fig:diststar}.}
\label{fig:distgal}
\end{figure*}

\subsection {Visual Inspection}
\label{eyeinspection}

The visual inspection of the catalogs was done using the new version
of ESO SkyCat which also provides the possibility of accessing the EIS
catalogs through the on-line server. Further information on the SkyCat
setup can be found at ``http://www.eso.org/eis/eis\_soft.html''.  This
setup interprets the parameters and flags available in the EIS
catalogs.  To distinguish between the different object classes and
flags, the following plot symbols and colors have been used:

\begin{itemize}

\item White (black) circles - stellarity index $\ge$ 0.75 and $f_s<$~4. 

\item Red circles - stellarity index $\ge$ 0.75 and $f_s$ $\ge$ 4.

\item Yellow ellipses - stellarity index  $<$ 0.75 and $f_s$ $<$ 4. 

\item Red crosses - stellarity index $< $ 0.75 and $f_s$ $\ge$ 4. 
\end{itemize}

This tool has been extensively used to fine-tune the configuration
parameters used by SExtractor and WeightWatcher as well as to inspect
the performance of the filtering of the catalogs (see
Section~\ref{sectderived}).  Users of the catalogs should be
aware of the following features:

\begin{itemize}

\item Many spurious objects are present near or within masked regions.
This is most noticeable along the frame border, affected by vignetting,
and along the bad pixel column. Real objects close to the masked
regions may also go undetected.

\item The largest cosmic rays can be classified as objects.

\item Spurious objects can be found in presence of very bright stars. 

\item The de-blending can fail when one of the objects in a
merged or very close pair, is much fainter than the other one (the
faint object is included as part of the brighter). Failures can also be
occur for close objects of the same brightness when the seeing
is bad or the PSF is elliptical.

\item High surface brightness galaxies fainter than $I \sim 21$ might be 
given a high value for the stellarity index (larger than 0.8 or even 0.9).

\end{itemize}

The visual inspection shows that, by adopting the filtering criteria
described in the previous section, most of the spurious objects are
appropriately removed.

\subsection{Uniformity of the Detections}

As a first check on the quality of the object catalogs produced by the
pipeline it is important to examine the uniformity of the detections
across the effective area (excluding the regions masked out) of the
EMMI-frame. This is shown in Figures~\ref{uniform_ew}
and~\ref{uniform_ns}, where the normalized average counts of stars and
galaxies as a function of the east-west (Figure~\ref{uniform_ew}) and
north-south position (Figure~\ref{uniform_ns}) on the chip are
displayed.  The upper panels show the star counts brighter than $I =
21$, which is the limiting magnitude for 
reliable classification in patch A as a whole.  The lower panels show
the galaxy counts to the same limiting magnitudes.

The overall uniformity of the detections at magnitudes $I \leq 21$ is
good. A small decrease in the number of stars is seen at the upper edge
of the chip and is almost compensated by an increase in the galaxy
counts. This behavior is likely to be due to misclassifications caused
by the increase in size of the PSF as shown in section~\ref{ksb}. 
Proper handling  of a variable PSF is currently being
implemented in SExtractor and should be available before the final
release of the EIS data.

\void{
As it can be seen the star counts show a decrease of about 5\% at the
right ($x \gsim 1700$) and upper ($y \gsim 1500$) edges of the EMMI
frame.  By contrast, the galaxy counts show an upturn in the same
region, by approximately the same amount which implies that the total
number of detected objects is constant over the frame at this limiting
magnitude. The observed effects in the star and galaxy counts can be
explained by the distortions of the images as discussed in
section~\ref{ksb}. That analysis showed that stars are smeared out at
the "upper right" corner of the EMMI frame. Therefore, it is not
surprising that in this region stars may have different (smaller)
stellarity index than elsewhere in the frame. To show this effect two
curves are shown in the figures for different assumed criteria for the
star/galaxy classification. As it can be seen adopting a smaller
stellarity index decreases somewhat the effect. This shows that the
proper choice of the stellarity index depends on the exact scientific
goals.} 

\begin{figure}
\resizebox{\hsize}{!}{\includegraphics{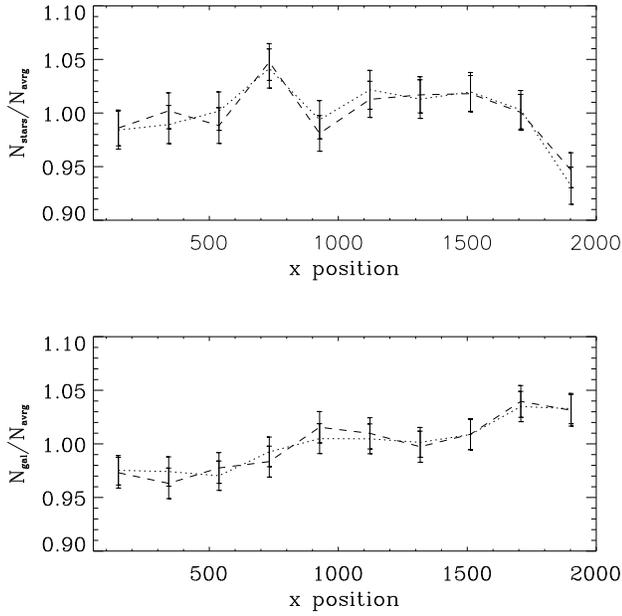}}
\caption{The uniformity of the detections in the east-west direction.
The top panel shows the detected stars brighter than $I=21$ for a
stellarity index $\ge$ 0.75 (dotted line) and stellarity index $\ge$ 0.5
(dashed line). For 
fainter magnitudes the classification breaks down. The bottom panel 
shows the detected galaxies brighter than $I=21$ stellarity index $<$
0.75 (dotted line) and stellarity index $<$ 0.5 (dashed line). It is
seen that the star counts show a dip at the "right" edge of the chip,
and a corresponding increase in the galaxy counts. This feature is
attributed to the image distortions, see text for details.}
\label{uniform_ew}
\end{figure}

\begin{figure}
\resizebox{\hsize}{!}{\includegraphics{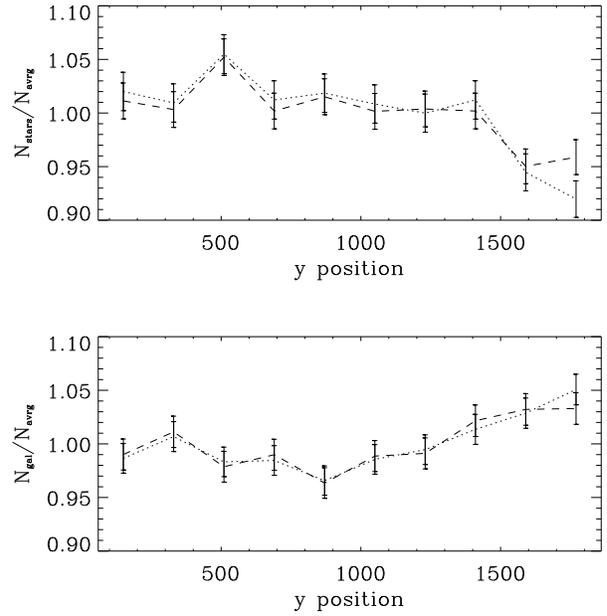}}
\caption{As figure ~\protect\ref{uniform_ew} but showing the detections 
in the north-south direction. Again at the "upper" edge of the chip we
see a dip in the star counts and a corresponding increase in galaxy
counts, which is due to the image distortions,see text for details.}
\label{uniform_ns}
\end{figure}

\subsection{Completeness and Reliability}

To verify the pointing of the telescope, a reference field has been
observed before the start of each row (150 sec) and, in some cases,
the start of sub-rows (50 sec). These exposures, which for patch~A
total 2250 sec, were used to determine the offset required to
compensate for the problems detected with the NTT pointing
model. Using the EIS pipeline these images have been coadded and an
object catalog produced extending to fainter magnitudes. This catalog
has been used to empirically determine the completeness of the
detections in typical single-frame EIS catalogs. This was done by
comparing the catalog produced from the coadded image of the reference
field to the the individual catalogs derived for the various exposures
of that field.  Since the limiting magnitude of the coadded image is
much fainter than that of the single frames, one can assume that the
coadded catalog is complete and that it is not significantly
contaminated with false objects at least to the limiting magnitude of
the single-frame catalogs.  Keeping this in mind a match was made
between all the objects in the coadded catalog and those found in the
single-frame catalogs. The ratio between the number of paired objects
and the total number of objects in the coadded catalog provides a
measure of the differential completeness as a function of magnitude,
which is shown in figure ~\ref{completeness}.  It is seen from the
figure that for objects of magnitude $I \sim 23 $ the completeness is
$\sim$ 80\%. At this magnitude the integrated completeness of the
catalog is 94\%. The completeness does not vary for seeing between
0.7 and 1.3 arcsec. For a seeing of 1.5 arcsec the 80\% differential
completeness limit is at $I \sim 22$.
 
\begin{figure}
\resizebox{\columnwidth}{!}{\includegraphics{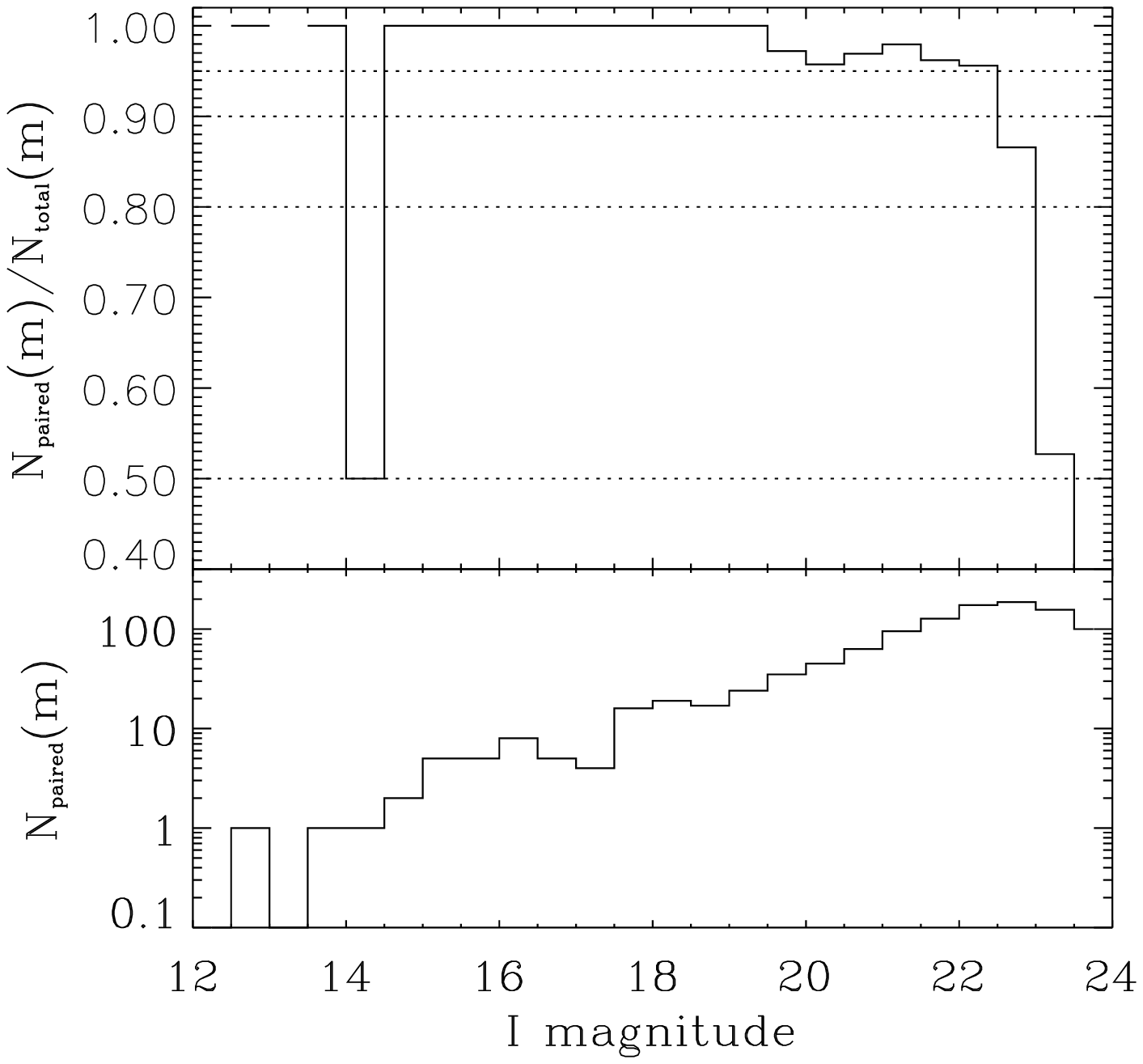}}
\caption{Top panel: Completeness of the single-frame catalogs as
determined from the comparison with the reference field. The plot shows
the ratio between the number of objects found in both the coadded image
and a single frame and objects found in the coadded image as a function
of magnitude. The figure shows that at $I \sim 23$ the single-frame
catalogs are 80\% complete. Bottom panel: The number of paired objects,
to give an idea of the statistical errors in the comparison.}
\label{completeness} \end{figure}

The number of false detections can also be estimated in the same
way. Figure~\ref{reliability} shows the ratio between objects that
were found in a single-frame catalog but not in the coadded one and
the total number of objects in the single-frame catalog.  The figure
is based on a comparison obtained using a single frame with a seeing
of 1.07 arcsec, which is close to the median seeing of the
observations for the patch.  It is seen that at $I \sim 22.5$ 10\% of
the objects are false detections and 20\% of the objects with
magnitude $I \sim 23$ are spurious. The integral fraction of spurious
objects up to the limiting magnitude of $I=23$ is $\sim 6\%$.

\void{
For the patch as a whole the 10 \% limit is
reached at $I \sim 21.5$ and 20\% at $I \sim 22$. As a reference, the
number counts for the catalog within the region analyzed is also
shown.}
 
\begin{figure}
\resizebox{\columnwidth}{!}{\includegraphics{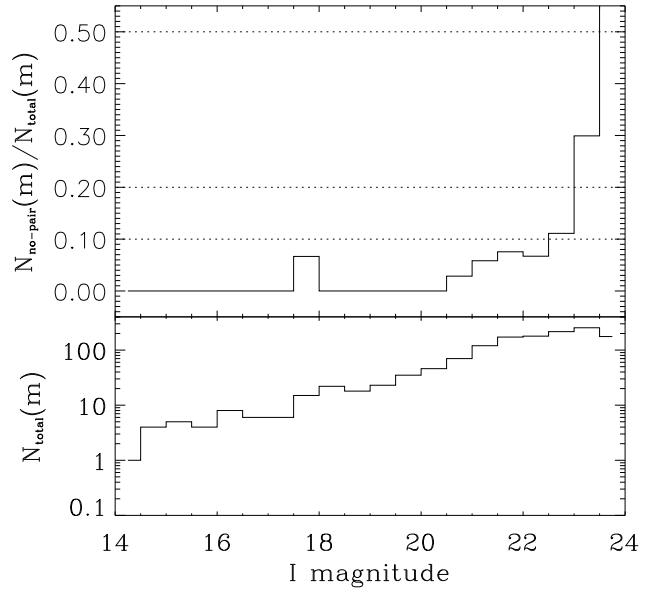}}
\caption{Top panel: ratio of the number of objects found in a single-frame but
not in the the coadded catalog of the reference field and total number
of objects in the corresponding single-frame catalog. At $I \sim 22.5$
there are 10 \% false objects and at $I \sim 23 $ there are 20\%
spurious detections. Bottom panel: total number counts in the
single-frame catalog.}
\label{reliability}
\end{figure}

\subsection{Errors in Magnitude and Classification}

A comparison between even and odd catalogs provides further useful
information on the accuracy of the magnitudes and on the robustness of
the classification as a function of magnitude. This comparison was
done using a test region of 0.6 square degrees, with a median seeing
of 0.95~arcsec. Using the same pairing procedure previously discussed,
a catalog of paired objects in the test region was produced.

A lower limit estimate of the photometric errors can be obtained from
the repeatability of the magnitudes of the paired
objects. Figure~\ref{mag_diff.ps} shows the magnitude difference of
these objects as a function of magnitude.  The standard deviation of
the magnitude differences in the interval $16 < I < 20.5$ ranges
between 0.02 and 0.1, reaching 0.3 at $I \sim 23$.
 
\begin{figure}
\resizebox{\columnwidth}{!}{\includegraphics{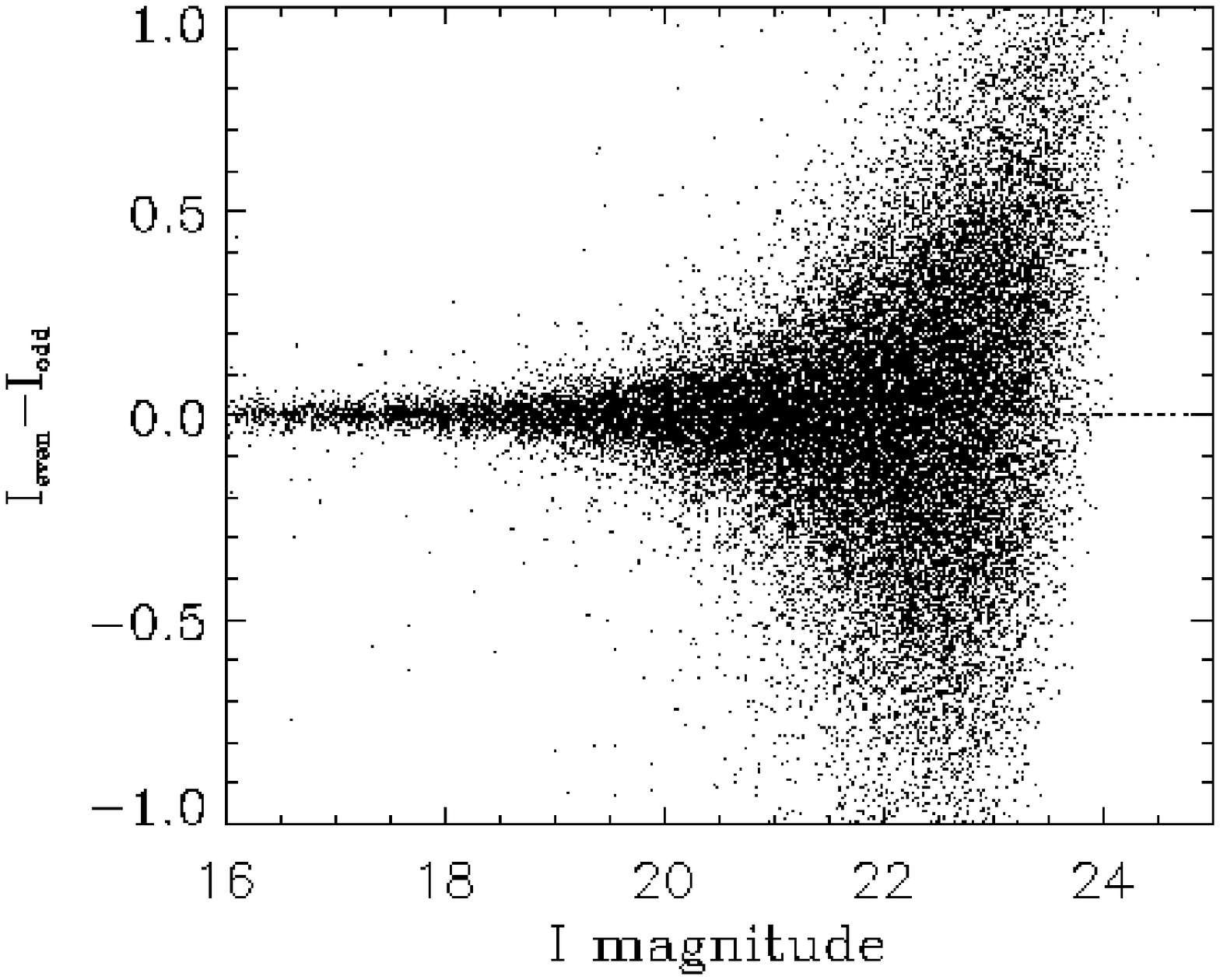}}
\caption{The magnitude differences between detections in the even and odd 
catalogs as function of magnitude. At bright magnitudes the standard deviation 
varies between 0.03 mag and 0.1 mag, reaching 0.4 mag at $I \sim 23$.}
\label{mag_diff.ps}
\end{figure}

Figure~\ref{mag_err} shows a comparison between the errors determined
from the magnitude difference shown above (divided by $\sqrt{2}$) and
the SExtractor error estimates based on photon statistics. SExtractor
provides reasonable error estimates over the interval of interest. At
bright magnitudes photometric errors are dominated by effects such as
flatfield errors, image quality, intrinsic stability of the MAG\_AUTO
estimator and relative photometry.

\begin{figure}
\resizebox{\columnwidth}{!}{\includegraphics{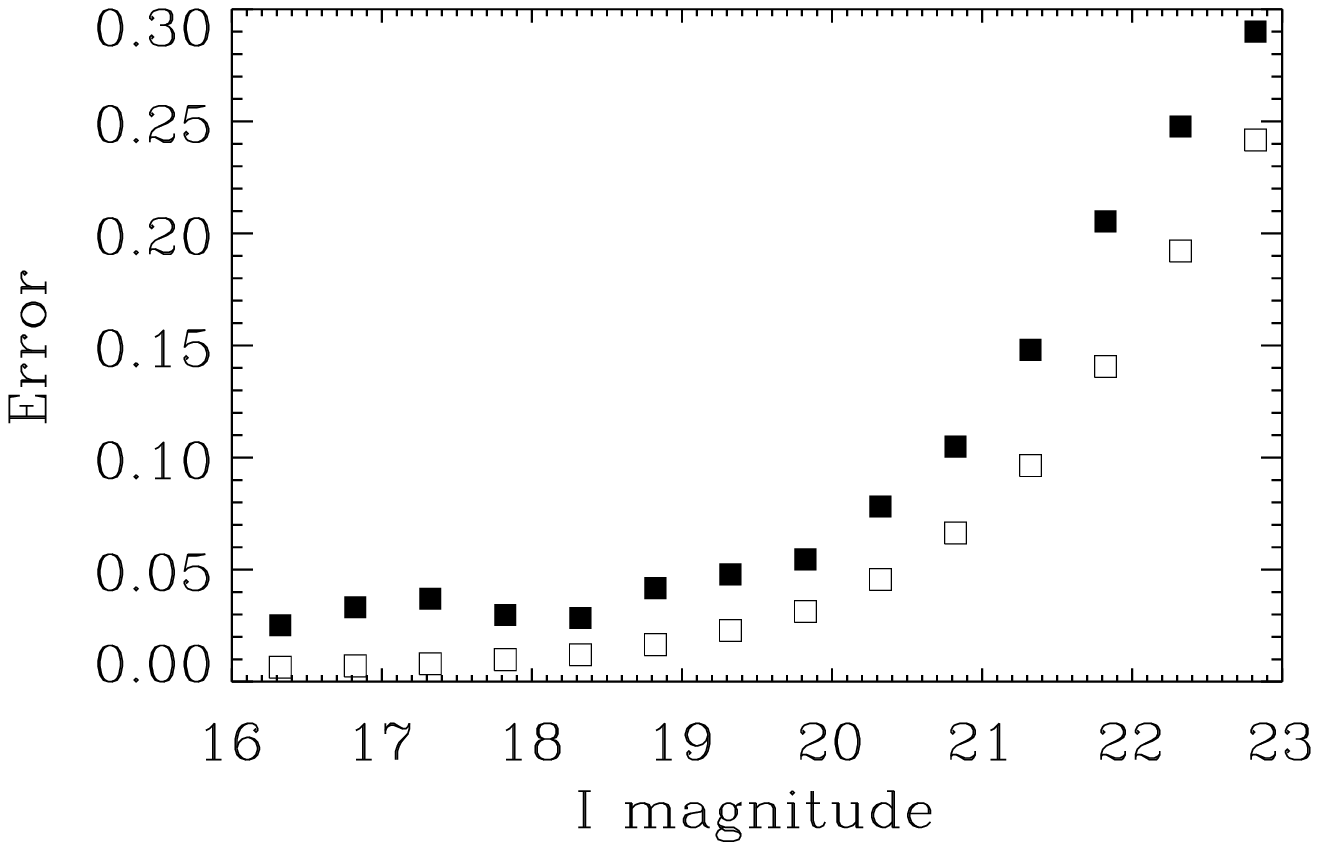}}
\caption{Comparison between the standard deviation of the
EIS-magnitude differences (filled squares) and the magnitude errors
estimated by SExtractor (open squares) as function of magnitude.}
\label{mag_err}
\end{figure}

For objects in the magnitude range $16 < I < 21$ and adopting a 
stellarity index of 0.75 to separate stars and galaxies, about 5\% of 
the objects 
have different classifications in the even and the odd catalogs.  
For magnitudes $I \lsim 16$ most objects are saturated and can be
classified as galaxies. However, they can be found as having the flag
$f_w=16$, which has been used to exclude them from subsequent analysis.

\section{Data evaluation}

\label{results}
\subsection{Galaxy and Stellar Counts}

Since EIS observations were carried out under varying conditions, it
is important to evaluate the degree of homogeneity of the final object
catalogs. This can be done by examining, for instance, the number
counts as a function of magnitude and comparing with earlier work.  To
evaluate the variation in the number counts due to the varying
observing conditions, the area of patch~A covered by both even and odd
tiles, comprising 3 square degrees, has been divided into six
subregions, each having an area of 0.5 square degrees. Note that these
areas cover most of patch~A including highly incomplete regions (see
Figure~\ref{fig:distgal}). The number counts for each subregion are
computed and the mean is shown in Figure~\ref{ncounts}, where the
error bars correspond to the standard deviation as measured from the
observed scatter in the six sub-catalogs. From the figure it can be
seen that the difference between the even and odd catalogs is
negligible.

\begin{figure}
\resizebox{9cm}{!}{\includegraphics{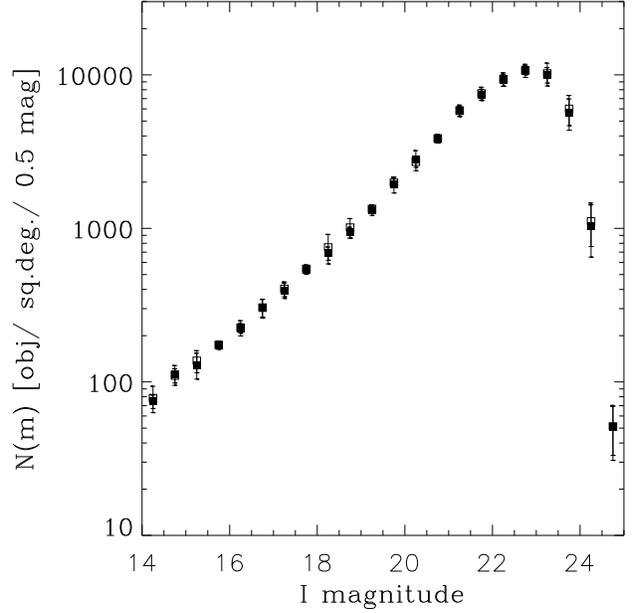}}
\caption{The object counts as a function of magnitude as derived from the average
of the counts in six odd (open squares) and even (filled squares)
sub-catalogs.  The error bars are the sample rms. In some cases the
error bars are of the same size as the symbols.}
\label{ncounts}
\end{figure}

In figure~\ref{ncounts_gal} the galaxy counts derived from the EIS
catalogs are compared to those of Lidman \& Peterson (1996) and
Postman \etal (1996). The $1 \sigma$ error bars are computed as above.
There is a remarkable agreement between the EIS galaxy-counts and
those of the other authors. The slope of the EIS counts is found to be
$0.43 \pm 0.01$. Also note that the EIS counts extend beyond those of
Postman \etal (1996) even for the counts derived from single
frames. The galaxies have been defined to be objects with a stellarity
index $<$ 0.75 for $I < 21$ and all objects fainter than $I =
21$. At this limit galaxies already outnumber stars by a factor of $\sim$ 3.

\begin{figure}
\resizebox{9cm}{!}{\includegraphics{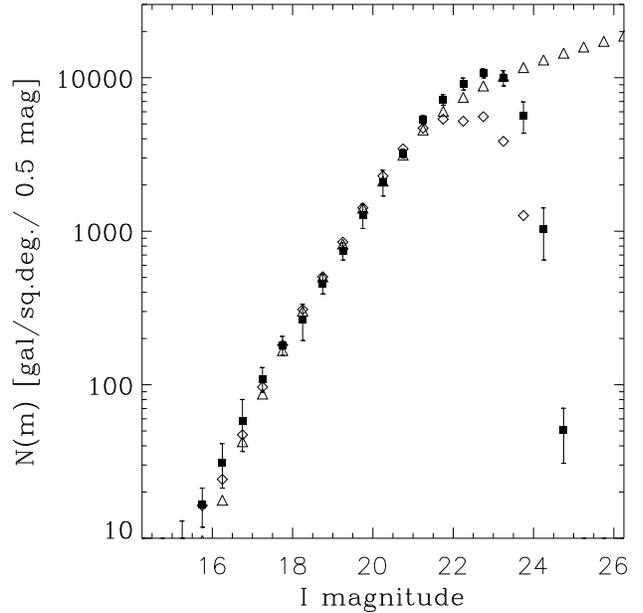}}
\caption{The EIS galaxy counts (filled squares) with $1\sigma$ error bars
compared to the galaxy counts derived by \protect\cite{lidman}
 (triangles) and
\protect\cite{postman} (diamond). The data from the other authors have
 been converted to the Johnson-Cousins system.}
\label{ncounts_gal}
\end{figure}

As discussed before, the criteria adopted for classifying stars and
galaxies is somewhat arbitrary. While a large value for the stellarity
index is desirable to extract a galaxy catalog as complete as
possible, this may not be the most appropriate choice for extracting
stellar samples. This can be seen in Figure~\ref{ncounts_star} where
the EIS star-counts are shown for two different choices of the
stellarity index for objects brighter than $I = 21$. For comparison,
the star counts predicted by the galactic model of M\'endez \& van
Altena (1996) are shown. As can be seen the observed counts agree with
the model for low values of the stellarity index (0.5), while higher
values shows a deficiency of stars at the faint-end.

\begin{figure}
\resizebox{9cm}{!}{\includegraphics{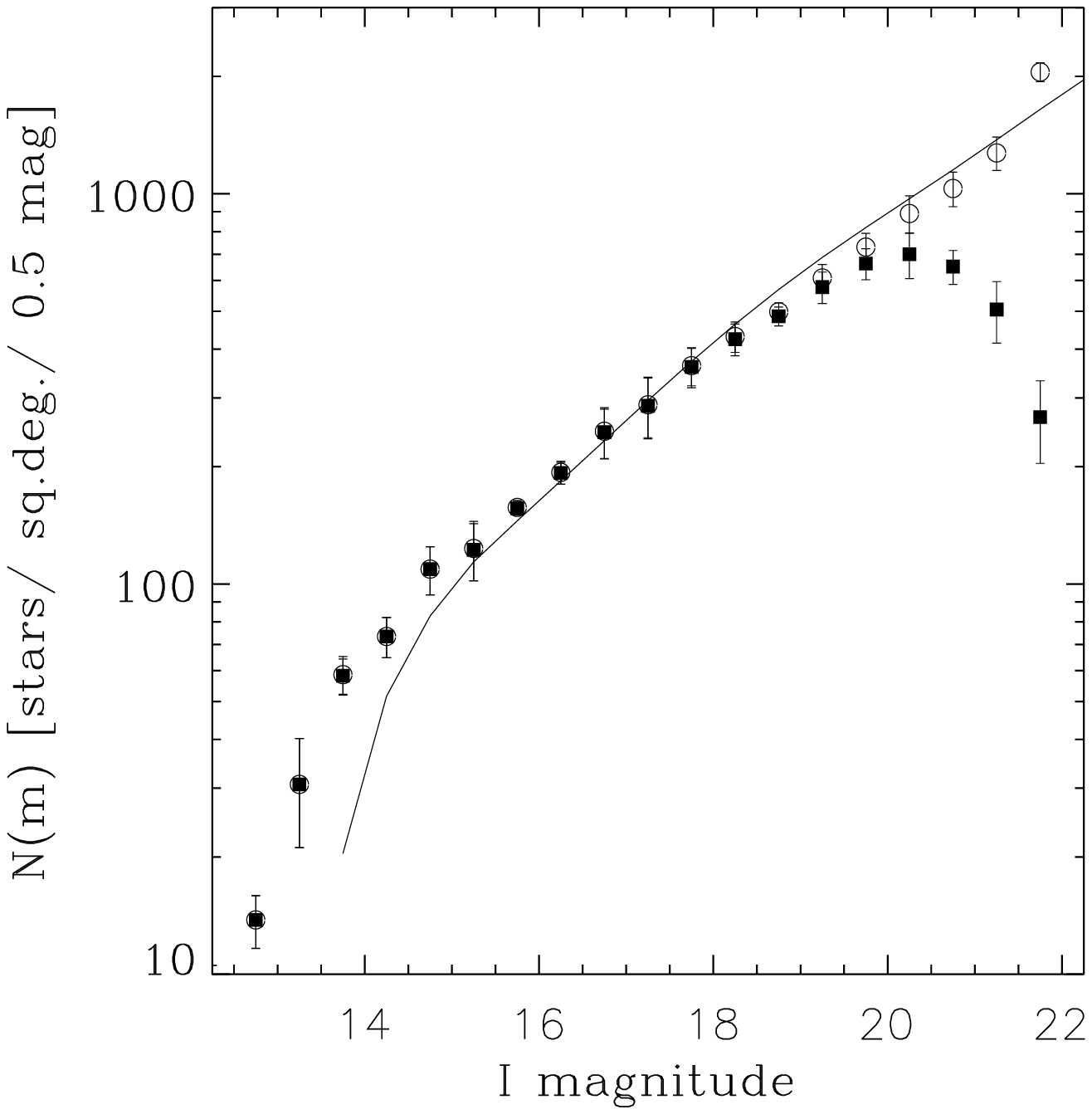}}
\caption{The EIS star counts  for stellarity index $\geq$ 0.75 (filled
squares) and for stellarity index $\geq$ 0.5 (open circles) compared
to the model by \protect\cite{mendez} (solid line).}
\label{ncounts_star}
\end{figure}

These preliminary results based on single-frame catalogs indicate that
the depth of the survey is close to that originally expected and is
sufficiently deep, especially after co-addition, to search
for distant clusters of galaxies, one of the main science goals of EIS.

\subsection {Angular Correlation Function}
\label{correlation}

In this section, the characteristics of the EIS catalogs are examined
by computing the angular correlation function over the whole patch,
for limiting magnitudes in the range $I=19$ to
$I=23$. Table~\ref{tab:angcorr} gives the number of galaxies down to
the different magnitude limits. For comparison the odd and even
catalogs are treated separately. The region defined by $\alpha \gsim
340^\circ$ and $\delta \gsim -39.8^\circ$ has been excluded from the
analysis because of its known incompleteness (section~\ref{field}).
Therefore, the total area used here is 2.38 square degrees.

\begin{table}
\caption{Number of Galaxies in Patch A}
\label{tab:angcorr}
\begin{tabular}{ccccc}
\hline
\hline
&\multicolumn{2}{c}{Even} &\multicolumn{2}{c}{Odd} \\
\hline
$m_{lim}$ & $N_{gal}$ &$\bar{n}$ & $N_{gal}$ &$\bar{n}$\\
          &           &$deg^{-2}$ &           &$deg^{-2}$\\
\hline
$m \le 19$ &2514   &1056.3   &2561    &1076.0 \\
$m \le 20$ &6972   &2929.4   &7072    &2971.4 \\
$m \le 21$ &18821  &7908.0   &19009   &7987.0 \\
$m \le 22$ &48129  &20222.3  &48316   &20300.8 \\
$m \le 23$ &94689  &39785.3  &95240   &40016.8 \\
\hline
\hline
\end{tabular}
\end{table}

To compute the angular correlation function, a random catalog is
created with the same geometry as the EIS catalog. The number of
random points has been chosen in order to yield an error less than
$10\%$ on the measured amplitude of $w(\theta)$ at $\theta=5$ arcsec.
The estimator used is that  described by Landy \& Szalay (1993):
\medskip
\begin{equation}
w(\theta)=\frac{DD-2DR+RR}{RR},
\end{equation}
\medskip
where $DD$, $DR$, and $RR$ are the number of data-data, data-random,
and random-random pairs at a given angular separation $\theta$.

In Figure~\ref{fig:w}, $w(\theta)$ for the even and odd catalogs are
compared. The error bars are 1$\sigma$ errors calculated with 10
bootstrap realizations. The angular correlation function  $w(\theta)$ is well
described by  a power-law $\theta^{-\gamma}$ with $\gamma
\sim 0.8$ (shown by the dashed line) over the entire range of angular
scales, extending out to $\theta \sim$ 0.5 degrees. In particular,
there is no evidence for any feature related to the scale of the EMMI
frame. Similar results are obtained when the angular correlation
function is computed from counts-in-cells.

\begin{figure}
%\resizebox{\height}{!}{\includegraphics{patchA_even_cor.ps}}
\resizebox{9cm}{!}{\includegraphics{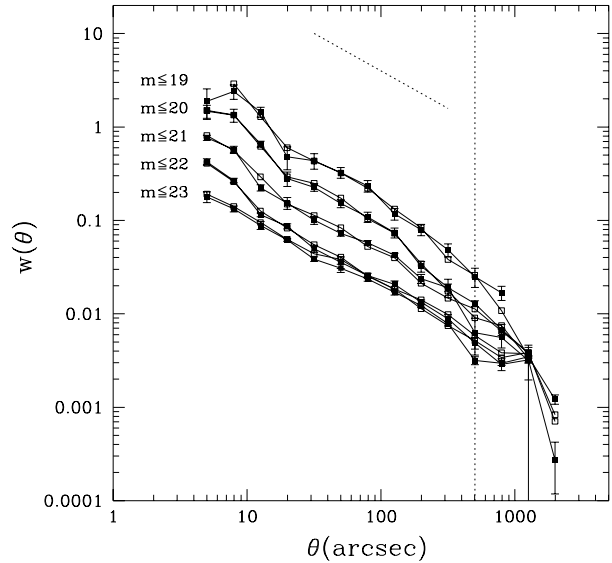}}
\caption[patchAevencor]{Angular two-point correlation function
calculated for the whole patch (even and odd catalogs) except the region 
defined by $\alpha \gsim 340^\circ$
and $\delta \gsim -39.8^\circ$ which has been removed for completeness
problems. The dotted line represents a power-law with slope of $-0.8$}
\label{fig:w}
\end{figure}

In Figure~\ref{fig:Aw}, the amplitude, $A_w(I)$, of the correlation
function at a scale of 1 arcsec is shown as a function of the limiting
magnitude. The amplitude was computed from the best linear-fits over
the range $\sim$ 10-200 arcsec of the $w(\theta)$, shown in
Figure~\ref{fig:w}, imposing a constant slope of $\gamma = 0.8$.  For
comparison, the power-law $A_w \propto 10^{-0.27 R}$, originally
determined by Brainerd \etal (1996) in the R-band is also shown
corrected for the mean color difference $(R-I)$ = 0.6 (Fukugita, Shimasaku \&
Ichikawa 1995).  The agreement with previous results
is excellent and demonstrates the good quality of the EIS catalogs,
even for a patch observed under less than ideal conditions.

\begin{figure}
\resizebox{9cm}{!}{\includegraphics{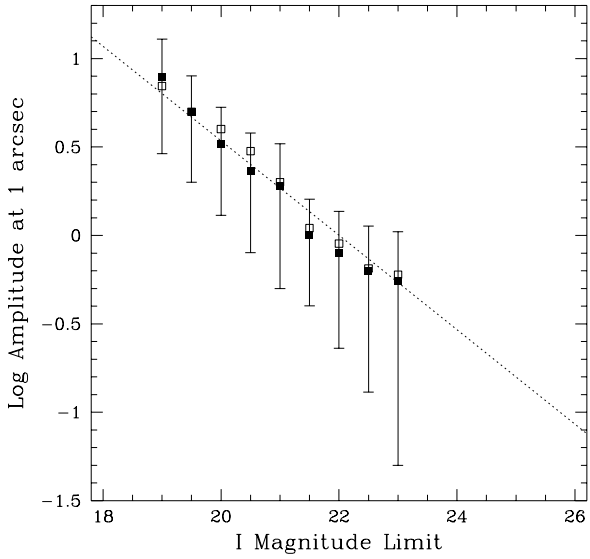}}
\caption[Aw]{Amplitude of the correlation function at 1 arcsec
calculated from the best fits of Figure~\ref{fig:w} as a function of limiting
magnitude. The dashed line is from from Brainerd \etal (1996).}
\label{fig:Aw}
\end{figure}

In order to evaluate the contribution of large-scale clustering to the
variance in the galaxy counts computed in the previous section, the
correlation function has been computed for the sub-areas used in
estimating the number counts.  This also offers the opportunity to
compare the amplitude of the cosmic variance to the bootstrap errors
used above.  The results are shown in Figure~\ref{fig:cosmicvar} where
the mean value over the sub-areas of $w(\theta)$ for each magnitude 
limit is presented. The error bars, that correspond to the rms of the six
sub-areas, are consistent with those found from the bootstrap
technique.

\begin{figure}
\resizebox{9cm}{!}{\includegraphics{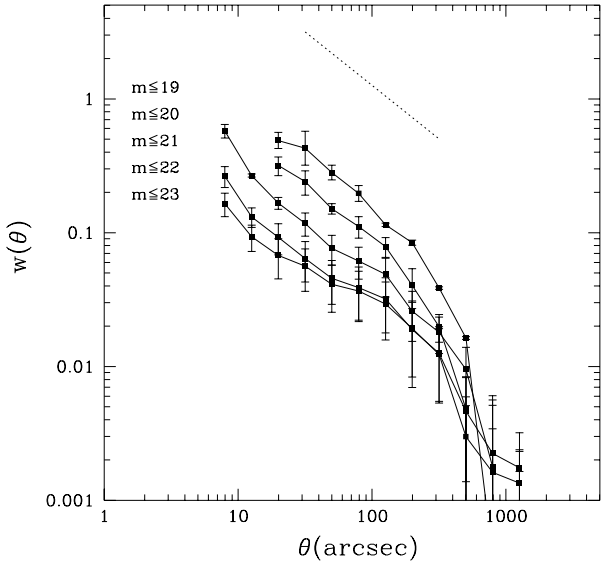}}
\caption[patchAcosmicvar6]{Mean value of the angular
correlation function measured for 6 sub-catalogs covering the whole
patch. The error bars are the standard deviations calculated between
these subsamples, and are therefore an estimate of the cosmic variance
on the corresponding scales.}
\label{fig:cosmicvar}
\end{figure}

\subsection{Field-to-field variations}
\label{field}

The data for patch A are far from homogeneous and for some
applications it is important to be able to objectively characterize
the area and location of regions not suitable for analysis.  A crude
selection of these regions can be done by examining the distribution
of the galaxy counts per frame as a function of the limiting
magnitude. This shown in Figure~\ref{fig:counts_dist} for the even and odd
frames. The vertical lines correspond to the 3$\sigma$ deviations from
the mean, where $\sigma$ includes the contribution from Poisson
fluctuations and the galaxy clustering. The latter was computed from
counts-in-cells of galaxies on the scale of an EMMI-frame in the best
available region of patch A.

\begin{figure}
  \resizebox{\columnwidth}{!}{\includegraphics{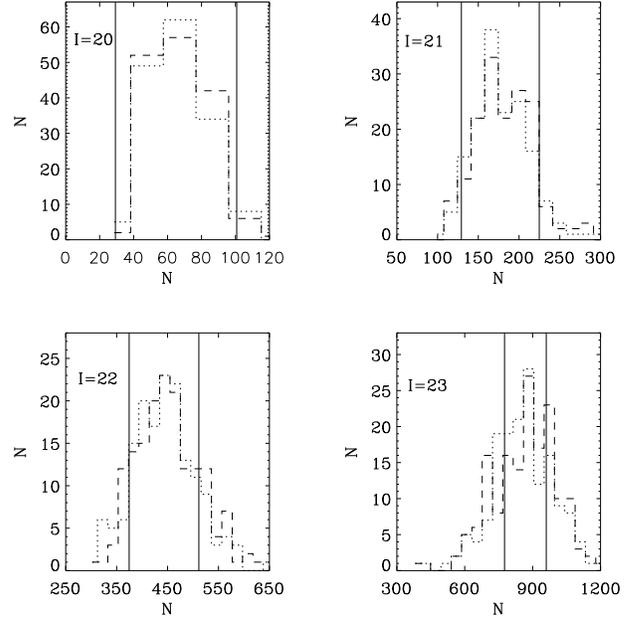}}
  \caption{Distribution of galaxy counts for the even (dotted line)
  and the odd (dashed line) catalogs as a function of limiting
  magnitude.  The vertical lines correspond to the 3$\sigma$
  deviations from the mean, where $\sigma$ includes the contribution
  from Poisson fluctuations and the galaxy clustering.}
  \label{fig:counts_dist}
\end{figure}

The results show that for galaxies brighter than $I = 20$, all frames
fall within the 3$\sigma$ level, except four, and the varying
observing conditions do not seem to affect the completeness of these
bright object catalogs. Going to fainter magnitudes the fraction of frames
with low number counts increases. This can be used to estimate the
size of the homogenous part of the patch at a given limiting magnitude
as shown in Figure~\ref{fig:bad_frac}. From the figure, it can be seen that
for $I \sim 22$ and $I \sim 23$ the useful area for analysis
corresponds to $\sim$90\% and $\sim$75\%, respectively. Most of the
rejected tiles are located in the region  $\alpha \gsim 340^\circ$ and
$\delta \gsim -40^\circ$ (Section~\ref{correlation}).

\begin{figure}
  \resizebox{\columnwidth}{!}{\includegraphics{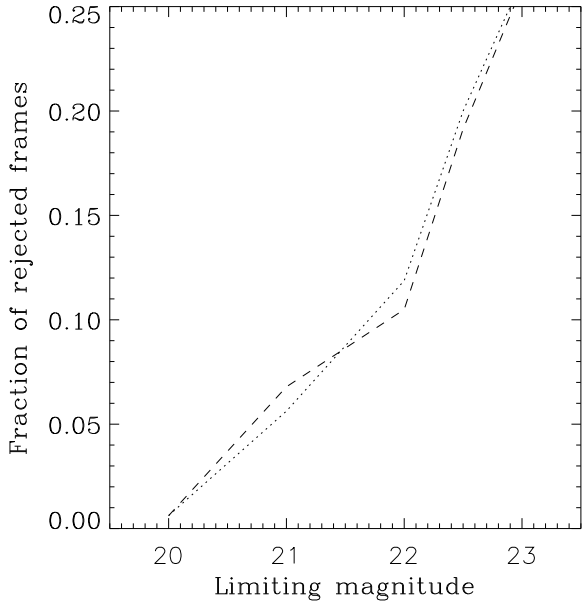}}
  \caption{Fraction of rejected tiles, corresponding to counts lower
  than 3$\sigma$ from the mean in Figure~\ref{fig:counts_dist}, as a
  function of the limiting magnitudes.}  \label{fig:bad_frac}
\end{figure}

\section{Future Prospects}

The full release of the EIS-wide data, except for the U-band
observations of patch B, is scheduled for July 1998. At that time, in
addition to the calibrated frames and single-frame catalogs the
coadded images and associated weight and context maps and catalogs
will become available. Information relative to the contexts, which for
coadded images replace the concept of a frame. is critical for the
suitable characterization of the data for analysis purposes as was
done in this paper on a frame basis.  Only with the context maps it
will be possible to extract sub-catalogs covering areas which fulfill
particular requirements of homogeneity.

In parallel, the EIS pipeline will be revised and upgraded to handle
data from CCD mosaics (SUSI2 and WFI@2.2), a key element for EIS-deep
and the Pilot Survey envisioned for the the first three months of
operation of the ESO/MPIA 2.2m telescope.  One of the goals of the
latter is to complete EIS-wide by providing multicolor coverage of
some of the EIS patches already observed in the I-band. Another
important task is to improve the data rate of the pipeline which will
be essential to cope with the expected data flow from the 2.2m
telescope.

Complementing the EIS pipeline, software has also been implemented to
produce derived catalogs which will also be made public.  Priority has
been given to searches of clusters of galaxies and preliminary results
based on the single frames of patch A will be reported in separate
papers of this series (Olsen \etal 1998, Slijkhuis \etal 1998).

All software and associated documentation will also be made available
together with an in-depth discussion of the algorithms used in the
pipeline (Deul \etal 1998). For the time being the upgraded SExtractor
can be found in the EIS home page.

\section{Summary}

The ESO Imaging Survey is being carried out to help the selection of
targets for the first year of operation of VLT. This paper describes
the motivation, field and filter selection, and data reduction
pipeline.  Data for the first completed patch, in the form of
astrometric and photometric calibrated pixels maps, single-frame
catalogs, on-line coadded section images and further information on
the project are available on the World Wide Web at
``http://\-www.eso.org/eis''.

Preliminary evaluation of the data shows that the overall quality of
is good and the completeness limit of the extracted catalogs is
sufficiently deep to meet the science requirements of
EIS. Furthermore, the results for the other patches should improve as
the observing conditions were considerably better than those in the
period patch A was observed.

The final and complete release of the data products of EIS is
scheduled as follows: 1) EIS-wide, except the U-band : July 31, 1998,
before the first call for proposals for the VLT; 2) EIS-deep and
EIS-wide U-band on December 31, 1998.

\begin{acknowledgements}

These data were taken at the New Technology Telescope (NTT) at the ESO 
La Silla Observatory under the ESO program identifications 59.A-9005(A) and 
60.A-9005(A). \\
We thank all the people directly or indirectly involved in the ESO
Imaging Survey effort. In particular, all the members of the EIS
Working Group for the innumerable suggestions and constructive
criticisms, the ESO Archive Group, in particular M. Albrecht, for
their support and for making available the computer facilities, ST-ECF
for allowing some members of its staff to contribute to this
enterprise. We also thank the Directors of all observatories and
institutes listed in this paper as affiliations for allowing the
participation of some of their staff in this project and for
suggesting some of their students and post-docs to apply to the EIS
visitor program. Special thanks to G. Miley, who facilitated the
participation of ED in the project and for helping us secure
observations from the Dutch 0.9m telescope.  To the Geneva
Observatory, in particular G. Burki, for monitoring the extinction
during most of the EIS observations.  To the NTT team for their help.
We are also grateful to N. Kaiser for the software and to the DENIS
consortium for making available some of their survey data. The DENIS
project development was made possible thanks to the contributions of a
number of researchers, engineers and technicians in various
institutes. The DENIS project is supported by the SCIENCE and Human
Capital and Mobility plans of the European Commision under the grants
CT920791 and CT940627, by the French Institut National des Sciences de
l'Univers, the Education Ministry and the Centre National de la
Recherche Scientifique, in Germany by the State of Baden-Wurttemberg,
in Spain by the DGICYT, in Italy by the Consiglio Nazionale delle
Richerche, by the Austrian Fonds zur F\"orderung der
wissenschaftlichen Forschung und Bundesministerium f\"ur Wissenschaft
und Forschung, in Brazil by the Fundation for the development of
Scientific Research of the State of S\~ao Paulo (FAPESP), and by the
Hungarian OTKA grants F-4239 and F-013990 and the ESO C \& EE grant
A-04-046. Special thanks to A. Baker, D. Clements, S. Cot\'e, E.
Huizinga and J. R\"onnback, former ESO fellows and visitors for their
contribution in the early phases of the EIS project. Our special
thanks to the efforts of A. Renzini, VLT Programme Scientist, for his
scientific input, support and dedication in making this project a
success. Finally, we would like to thank ESO's Director General
Riccardo Giacconi for making this effort possible.

\end{acknowledgements}


\begin{thebibliography}{}

\bibitem[Baldwin \& Stone 1984]{baldwin}
Baldwin, J.A., Stone, R.P.S., 1984, MNRAS 206, 241
\bibitem[Bertin 1998]{bertin:1998}
Bertin, E., 1998, {\it in preparation}
\bibitem[Bertin \& Arnouts 1996]{bertin}
Bertin, E. \& Arnouts, S., 1996, A\&AS 117, 393
\bibitem[]{} Brighton, A., 1998, VLT-MAN-ESO-19400-1552, Issue 1.0, Version 2.0.4, "The ESO
SkyCat Tool Astronomical Image and Catalog Browser Programmer's
Manual"
\bibitem[]{} Brainerd, T.G., Smail, I.R., Mould, J.R., 1995, MNRAS 275, 781
\bibitem[da Costa et al. ??]{dacosta}
da Costa \etal 1998a, The Messenger, 91, 49
\bibitem[da Costa et al.]{dacosta}
da Costa \etal 1998b, {\it in preparation}
\bibitem[deul]{} Deul \etal 1998,  {\it in preparation}
\bibitem[]{} Devillard N., 1997, http://www.eso.org/eclipse
\bibitem[]{} D'Odorico, S., 1990, The Messenger, 61, 51
\bibitem[Epchtein et al. 1996]{denis}
Epchtein, N., et al., 1996, The Messenger 87, 27
\bibitem[]{} Epchtein, N., 1997, Proceedings of the  Euroconference on
``Impact of Large Scale Near-Infrared Surveys'', eds. F. Garzan,
N. Epchtein, A. Omont, W. B. Burton, P. Persi (Kluwer Academic
Publishers:Dordrecht)
\bibitem[]{} Erben, T. 1996, {\it private communication}
\bibitem[Fruchter \& Hook 1997]{fruchter}
Fruchter, A.S. \& Hook, R.N., 1997, in Applications of Digital Image 
Processing XX, ed. A. Tescher, Proc. S.P.I.E. vol. 3164, 120
\bibitem[]{} Fukugita, M., Shimasaku, K., \& Ichikawa, T., 1995, PASP 107, 945
\bibitem[]{} Greisen E.W. \& Calabretta M., 1996, \\ http://www.cv.nrao.edu/fits/documents/wcs/wcs.html
\bibitem[Gunn \& Stryker 1983]{gunn}
Gunn, J.E. \& Stryker, L.L., 1983, ApJS 52, 121
\bibitem[Hook \& Fruchter 1997]{hook}
Hook, R.N. \& Fruchter, A.S., 1997, in ASP Conf. Series, Vol. 125, Astronomical Data Analysis 
Software and Systems VI, ed. G. Hunt and H.E. Payne (San Francisco: ASP), 147
\bibitem[Irwin 1985]{irwin}
Irwin, M.J., 1985, MNRAS 214, 575
\bibitem[Kaiser et al. 1995]{kaiser}
Kaiser, N., Squires, G., Broadhurst, T., 1995, ApJ 449, 460
\bibitem[Kron 1980]{kron}
Kron, R.G., 1980, ApJS 43, 305 
\bibitem[Landolt 1992a]{landolt1}
Landolt A.U., 1992a, AJ, 104, 340
\bibitem[Landolt 1992b]{landolt2}
Landolt A.U., 1992b, AJ, 104, 372
\bibitem[Landy 1993]{landy} 
Landy, S.D., \& Szalay, A.S., 1993, ApJ 412, 64.
\bibitem[Lidman \& Peterson 1996]{lidman} 
Lidman, C. \& Peterson, B., 1996, MNRAS 279, 1357
\bibitem[M\`{e}ndez \& van Altena 1996]{mendez} 
M\`{e}ndez, R. \& van Altena, W., 1996, AJ 112, 655
\bibitem[]{} Olsen \etal 1998, A\&A {\it submitted}
\bibitem[Postman et al. 1996]{postman} 
Postman, M., Lubin, L.M., Gunn, J.E., Oke, J.B., Hoessel, J.G., Schneider, 
D.P., Christensen, J.A., 1996, AJ 111, 615
\bibitem[]{} Renzini, A. \& da Costa, L.N., 1997,  The Messenger 87, 23
\bibitem[]{} Renzini, A. 1998, The Messenger, 91, 54
\bibitem[]{} Slijkhuis \etal 1998, {\it in preparation} 
\bibitem[]{} Villumsen \etal 1998 private communication
\bibitem[]{} White, R., 1992, http://www.stsci.edu/software/hcompress.html 


\end{thebibliography}
\end{document}